\documentstyle[12pt,epsf]{article}
\newcommand{\Kbar}{\not{\!K}}

\newcommand{\Hslash}{\not{\!H}}
\newcommand{\Qbar}{\not{\!Q}}

\newcommand{\Pbar}{\not{\!P}}

\newcommand{\be}{\begin{equation}}
\newcommand{\ee}{\end{equation}}
\newcommand{\ba}{\begin{eqnarray}}
\newcommand{\ea}{\end{eqnarray}}

\newcommand{\nh}{{\bf      h}}

\newcommand{\nk}{{\bf      k}}
\newcommand{\np}{{\bf      p}}       
\newcommand{\nq}{{\bf      q}}

\begin{document}
\begin{titlepage}
\mbox{} 
\vspace*{2.5\fill} 
{\Large\bf 
\begin{center}
%
Relativistic pionic effects in quasielastic electron scattering
%
\end{center}
} 
\vspace{1\fill} 
\begin{center}
{\large 
J.E. Amaro$    ^{1}$, 
M.B. Barbaro$  ^{2}$, 
J.A. Caballero$^{3,4}$, 
T.W. Donnelly$ ^{5}$ and 
A. Molinari$   ^{2}$
}
\end{center}
\begin{small}
\begin{center}
$^{1}${\sl 
Departamento de F\'\i sica Moderna,
Universidad de Granada, 
E-18071 Granada, SPAIN 
}\\[2mm]
$^{2}${\sl 
Dipartimento di Fisica Teorica,
Universit\`a di Torino and
INFN, Sezione di Torino \\
Via P. Giuria 1, 10125 Torino, ITALY 
}\\[2mm]
$^{3}${\sl 
Departamento de F\'\i sica At\'omica, Molecular y Nuclear \\ 
Universidad de Sevilla, Apdo. 1065, E-41080 Sevilla, SPAIN 
}\\[2mm]
$^{4}${\sl 
Instituto de Estructura de la Materia, CSIC 
Serrano 123, E-28006 Madrid, SPAIN 
}\\[2mm]
$^{5}${\sl 
Center for Theoretical Physics, Laboratory for Nuclear Science 
and Department of Physics\\
Massachusetts Institute of Technology,
Cambridge, MA 02139, USA 
}
\end{center}
\end{small}

\kern 1. cm \hrule \kern 3mm 

\begin{small}
\noindent
{\bf Abstract} 
\vspace{3mm} 

The impact of relativistic pionic correlations and meson-exchange
currents on the response functions for electromagnetic quasielastic
electron scattering from nuclei is studied in detail. Results in
first-order perturbation theory are obtained for one-particle emission
electronuclear reactions within the context of the relativistic Fermi
gas model. Improving upon previous analyses where non-relativistic
reductions of the currents were performed, here a fully relativistic
analysis in which both forces and currents are treated consistently is
presented.  Lorentz covariance is shown to play a crucial role in
enforcing the gauge invariance of the theory. Effects stemming
uniquely from relativity in the pionic correlations are identified
and, in particular, a comprehensive study of the self-energy
contributions and of the currents associated with the pion is
presented.  First- and second-kind scaling for high momentum transfer
is investigated.

\kern 2mm 


\end{small}
\kern 2mm \hrule \kern 1cm
\noindent MIT/CTP\#3127 \hfill May 31, 2001
\end{titlepage}

\section{Introduction}
\label{sec:Intro}


For years electron scattering has provided one of the most powerful means to
study the structure of nuclei. In particular, inclusive
(e,e$'$) processes at or near quasielastic peak kinematics have attracted
attention in the last two decades and a 
number of experiments have been performed with the aim of disentangling the 
longitudinal and transverse contributions to the quasielastic cross section.
Calculations based on a simple picture of the dynamics of the quasielastic
peak (one-photon-exchange, impulse approximation 
and one-body currents) appeared to be
successful in explaining the early (e,e$'$) cross section data for 
several nuclei~\cite{Mon71}, except at large energy transfer
where $\Delta$ excitation of the nucleon contributes substantially. 
However, a simultaneous explanation of the separated longitudinal, 
$R^L$,  and transverse, $R^T$, response functions could not be fully reached
when representing the quasifree peak simply as an incoherent 
sum of elastic nucleon scattering processes. Some authors attributed
this problem to a large quenching thought to occur in
$R^L$~\cite{Mez84,Alt80,Bar83,Orl91}, although others found no
strong evidence for such a 
quenching~\cite{Deady86,Williamson97,Blatchley86} and later
analysis of the world data~\cite{Jourdan} showed that 
the reduction of the longitudinal response at high momentum 
transfers was actually smaller than previously assumed. 
Thus at present it appears more likely that  
the ``quenching'' of the longitudinal
response is not where the problem lies, but rather that the transverse
response has contributions other than those mentioned above --- some
of these contributions constitute the focus of the present work.

Different approaches to this problem have been invoked. As mentioned 
above, the earlier approaches used only one-body (nucleonic) 
currents, {\it i.e.,}
the impulse approximation (IA). Whereas the IA explains the 
electron scattering reaction mechanism around the quasielastic 
peak reasonably well, theoretical models based on one-body currents
are unable to account for the observed strength in the dip 
region between the quasielastic and $\Delta$ peaks, where meson 
production, including via the $\Delta$, and two-body
currents~\cite{MECpapers} should be taken into account in the analysis. 
Several studies dealing with the issue of two-body currents
in quasielastic electron scattering reactions have been published. 
Most focused on the role played by the meson-exchange currents
(MEC), specifically the seagull and pion-in-flight terms. 
For instance, detailed results were obtained 
with the Fermi gas model (RFG)~\cite{VanOD81,ADM90,Dekker} and, 
as well, using a continuum shell model~\cite{Ama92,Ama94b,Ama94}. 
Moreover, the contribution of the $\Delta$
isobar current, {\it i.e.,} the current associated with the electromagnetic 
excitation of an intermediate isobar, which decays by exchanging a 
virtual pion with 
a nucleon, has been also considered \cite{ADM90,Ama94,Koh81}
and shown, in most cases, to be larger than the purely $\pi N$ 
MEC contributions.
 
In some previous work~\cite{Blu89} a relativistic analysis of the
MEC has been presented (although there the associated correlations
were neglected and thus gauge invariance was not respected); however,  
most calculations have been done basically within a non-relativistic
framework. As a consequence, not only have non-relativistic wave functions
been used, but also non-relativistic current operators, the latter 
being obtained using standard expansions which view both 
the dimensionless momentum transfer $\kappa\equiv q/2m$ and
the dimensionless energy transfer $\lambda\equiv \omega/2m$ as
being small ($m$ is the nucleon mass).
Obviously, for high-energy conditions, the current operators 
so obtained are inadequate and new expressions are needed.
This provided the focus for the developments in~\cite{Ama98} where new
current operators that are exact as far as the variables
$\omega$ and $q$ are concerned were derived.

As mentioned above, another difficulty of most of the non-relativistic
pionic calculations existing in the literature relates to the
breakdown of gauge invariance. In this connection, we recall some
differences between the pionic and isobar two-body current
operators. In a representation of the nucleus with states described
only in terms of nucleons and where current and interaction effects
are embodied in operators that contain all other hadronic degrees of
freedom (pions, deltas, ...) these two enter somewhat differently. In
lowest order the latter is gauge invariant by itself, whereas the
former enters both in one-boson-exchange interactions ({\it i.e.,}
potentials in the non-relativistic limit) and in one-boson-exchange
currents. Thus the two-body pionic currents must, via the continuity
equation, be connected with the interaction effects, the so-called
correlations. One might expect such pionic contributions to be more
constrained than those arising from the isobar. But in most
non-relativistic calculations the contribution of $\pi$ meson-exchange
currents violates current conservation. This constitutes a fundamental
flaw for the consistency of the theory and hence is one of the main
issues addressed in this work.

In this paper our aim is to investigate the role of pions in inclusive
electron scattering from nuclei within the kinematical regime of the
quasielastic peak. We restrict our attention to the study of the
purely electromagnetic response functions, $R^L$ and $R^T$, leaving
for subsequent presentation the analysis of parity-violating responses
and of the roles played in the responses by mesons heavier than the
pion.  Furthermore, here our focus is placed on the contributions that
affect the one-particle one-hole (1p-1h) sector of the nuclear
excitations, whereas in the near future we shall also present results
within the same framework for contributions that affect the 2p-2h
excitations.

This work extends the approach presented in a previous
paper~\cite{ADM90}, but introduces some important new elements, not
only in connection with the implementation of Lorentz covariance, but
also as far as the self-consistency of the theory (specifically the
fulfillment of gauge invariance) is concerned.  As in~\cite{ADM90} we
undertake this study within the context of the RFG model, which is
well suited for maintaining the consistency in the treatment of the
forces and currents while fullfilling the Lorentz covariance.  The RFG
is in fact the simplest model that provides a valid starting framework
for describing quasielastic electron scattering and exactly respects
relativity. To employ this model and to extend it using standard
perturbation theory to account for first-order pionic effects amounts
to considering all Feynman diagrams carrying one pionic line (a
discussion of the relevance of higher-order pionic diagrams can be
found in~\cite{Barb93,Barb94}). Accordingly, our theory is based on
these diagrams and we do not make any non-relativistic expansion when
handling the pionic current operators; herein lies the main difference
with previous work~\cite{ADM90,Ama98,Ama99} where some kind of
expansion in powers of $1/m$ was introduced.  In the present study we
present a fully relativistic analysis including the pionic correlation
effects, embodied in the so-called self-energy and exchange diagrams,
as well as the MEC contributions (seagull and pion-in-flight terms).
The analysis of the $\Delta$ current, within the same framework, will
be presented in a forthcoming publication.

With regard to gauge invariance, in contrast to previous
(non-relativistic) analyses, our findings strongly support the need
for a fully relativistic treatment in order to preserve this symmetry.
In addition, relativity is clearly manifested in new effects which
appear at the level of the currents and/or propagators and which are
absent in the non-relativistic approximations.  The relativistic
analysis of the self-energy (s.e.) term that presents some important
differences compared with the non-relativistic approach is
particularly significant.  This issue will be treated in detail in
forthcoming work~\cite{Ama01} and so here we only mention the fact
that two different contributions in the s.e. pionic correlation term
can be identified. One stems from the renormalization of the energy
and the other has its origin in the lower components of the Dirac wave
functions.  Whereas the first term can be shown to correspond to the
one already embodied in the s.e. diagram within the non-relativistic
approach~\cite{ADM90}, the second one has no counterpart in the
non-relativistic analysis.  However, both contributions are necessary,
not only to set up a consistent and well-defined self-energy
particle-hole (ph) current matrix element, but also to fulfill the
continuity equation expressing the gauge invariance of the theory.

This paper is organized as follows: in Section~\ref{sec:Form} we
present the general formalism and the basic ingredients that enter in
analyses of inclusive quasielastic electron scattering. We define the
longitudinal and transverse electromagnetic inclusive response
functions in terms of the hadronic tensor $W^{\mu\nu}$. This in turn
can be constructed through the ph current matrix elements or,
equivalently, as the imaginary part of the polarization propagator.
Section~\ref{sec:Pion} deals with the one-pion-exchange contributions
to the nuclear responses.  We start from the Feynman diagrams
corresponding to the free-space MEC and pion induced nucleon-nucleon
correlation currents.  We prove, at this level, the gauge invariance
of the theory. From these currents we next set up the ph matrix
elements that enter in our RFG-based computation.  The case of the
s.e. diagram is discussed in detail as the corresponding ph current
matrix element is divergent. We treat this term using the polarization
propagator approach, thus obtaining a finite response. In~\cite{Ama01}
we shall prove, however, that it is possible to derive a new
``renormalized'' expression for the self-energy current matrix element
which is finite, consistent with the analysis performed with the
polarization propagator technique and, when taken together with the
remaining pionic currents and the one-body current, satisfies current
conservation.  In Section~\ref{sec:tensor} we discuss the effects
introduced by the MEC and pion correlations at the level of the two
response functions, $R^L$ and $R^T$.  We study the magnitude of these
effects as a function of the transfer momentum $q$, and, as well,
their dependence upon the Fermi momentum $k_F$.  Such dependences are
clearly relevant in connection with analyses of scaling in
nuclei~\cite{yscaling}, both of first and second kinds, and, as we
shall see, scaling of the former kind is attained at high $q$, whereas
scaling of the latter kind is clearly broken through dependence
roughly on the Fermi momentum squared for all values of $q$.  The
issue of the impact on the responses of the dynamical pion propagator,
versus the standard static propagator, is also discussed.  Finally, in
Sections~\ref{sec:res} and \ref{sec:concl} we summarize our results
and draw our conclusions.

\section{General Formalism}
\label{sec:Form}

In this section we briefly summarize the general formalism involved in
the description of (e,e$'$) processes for quasielastic kinematics. We
only discuss those aspects that are of special relevance to the
analysis that follows and the reader who is interested in more details
will benefit from reading~\cite{Form1,Don92,Form2,Rassegna,Alv01}.  We
limit our attention to the Plane Wave Born Approximation (PWBA) in
which the electron is described as a plane wave and interacts with the
nuclear target via the exchange of a virtual photon.  The variables
involved in the process are $K^\mu_e=(\varepsilon,\nk)$ and
$K'^\mu_e=(\varepsilon',\nk')$, the initial and scattered electron
four-momenta, and $P_i^\mu=(M_i,{\bf 0})$ and $P_f^\mu=(E_f,\np_f)$,
the initial and final hadronic four-momenta, respectively.  The
four-momentum transferred by the virtual photon is
$Q^\mu=(K_e-K'_e)^\mu=(P_f-P_i)^\mu=(\omega ,\nq)$.  Assuming Lorentz
invariance and parity conservation, the inclusive (e,e$'$) cross
section then reads

\begin{equation}
\frac{d\sigma}{d\Omega'_e d\omega}=
\frac{2\alpha^2}{Q^4}\left(\frac{\varepsilon'}{\varepsilon}\right)
\eta_{\mu\nu}W^{\mu\nu}=
\sigma_M\left[v_L R^L(q,\omega) + v_T R^T(q,\omega)\right] \ ,
\label{eq1}
\end{equation}
where $\alpha$ is the fine structure constant, $\sigma_M$ the Mott
cross section and $\Omega'_e$ the scattered electron solid angle.  The
terms $\eta_{\mu\nu}$ and $W^{\mu\nu}$ represent the leptonic and
hadronic tensors, respectively.  The kinematic factors $v_L$ and $v_T$
are evaluated from the leptonic tensor with standard techniques
(see~\cite{Form1} for explicit expressions), whereas the longitudinal
and transverse (with respect to the momentum transfer $\nq$) response
functions $R^L$ and $R^T$ are constructed directly as components of
the hadronic tensor $W^{\mu\nu}$ according to
\begin{eqnarray}
R^L(q,\omega)&=&\left(\frac{q^2}{Q^2}\right)^2\left[
W^{00}-\frac{\omega}{q}(W^{03}+W^{30})+\frac{\omega^2}{q^2}W^{33}
\right]\label{eq2a} \\
R^T(q,\omega)&=&W^{11}+W^{22} \ .
\label{eq2b}
\end{eqnarray}
Note that if gauge invariance is fulfilled, implying that
$W^{03}=W^{30}=(\omega/q)W^{00}$ and $W^{33}= (\omega/q)^2 W^{00}$,
then $R^L$ is simply the time component of the hadronic tensor, namely
$W^{00}$.  Hence $R^L$ is determined by the charge distribution,
whereas $R^T$ reflects the current distribution of the nuclear target.

The hadronic tensor and consequently the response functions derived
from it embody the entire dependence on the nuclear structure,
specifically on the charge and current distributions in nuclei, and
accordingly these provide the prime focus in analyses of electron
scattering.  There are various options on how to proceed in performing
such analyses (see, for example, \cite{Chinn}), depending on the
specific problem under consideration and on the approximations to be
made. In what follows we recall two common expressions for the
hadronic tensor $W^{\mu\nu}$ and comment briefly on their
applications.

First, the hadronic tensor can be defined according to
\begin{equation}
W^{\mu\nu}=\overline{\sum_i}\sum_f\langle f|\hat{J}^\mu |i\rangle^\ast
\langle f|\hat{J}^\nu |i\rangle \delta(E_i+\omega-E_f) \ ,
\label{eq3}
\end{equation}
where $\hat{J}^\mu$ represents the nuclear many-body current operator,
the nuclear states $|i\rangle$ and $|f\rangle$ are exact
eigenstates of the nuclear Hamiltonian with definite four-momentum,
and the sum with a bar means average over initial states.
This form is very general and includes all possible final states that
can be reached through the action of the current operator
$\hat{J}^\mu$ on the exact ground state.  In our perturbative approach
we shall use eigenstates of the free Hamiltonian $H_0$ (which
describes the free RFG) and include correlations among nucleons in the
current mediated by the exchange of pions (MEC).  This current of
course allows one to reach both the p-h and the 2p-2h sectors in the
Hilbert space of $H_0$. In the present work, however, we shall
restrict our attention to the former.

A different option for evaluating the nuclear responses 
exploits the polarization propagator $\Pi^{\mu\nu}$ 
(also referred to as the current-current
correlation function). The latter can be expressed in terms of the
full propagator, $\hat{G}$, of the nuclear many-body system, since closure
can be used to carry out the sum over
the final states in eq.~(\ref{eq3}). Then one has for 
the hadronic tensor
\begin{equation}
W^{\mu\nu}=
-\frac{1}{\pi} {\rm Im}\, \left(\Pi^{\mu\nu}\right)
=-\frac{1}{\pi} {\rm Im}\,
\left(\overline{\sum_i}\langle i|\hat{J}^{\dagger\mu}
\hat{G}(\omega +E_i)\hat{J}^\nu |i\rangle\right) \ .
\label{eq4}
\end{equation}
A possible advantage of this approach relates to the existence of a 
well-defined set of rules (the relativistic Feynman diagrams) which 
allows one to compute
$\Pi^{\mu\nu}$ perturbatively~\cite{FW}.

Obviously both procedures are equivalent and hence the observables
calculated using the expressions for the hadronic tensor given 
by eqs.~(\ref{eq3}) or (\ref{eq4}) should be the same. However, notice
that eq.~(\ref{eq3}) is less suitable for dealing with situations where
the nuclear current matrix element $\langle f|\hat{J}^\mu |i\rangle$
is divergent. In this case it is advisable to compute the responses via the
polarization propagator.

Finally, we remark that gauge invariance must be fulfilled both at the
level of the nuclear current matrix elements and at the level of the
hadronic tensor and/or the polarization propagator. A consequence is
that the electromagnetic continuity equation should be satisfied.  In
other words in momentum space all of the expressions $Q_\mu\langle
f|\hat{J}^\mu |i\rangle$, $Q_\mu W^{\mu\nu}$ and $Q_\mu\Pi^{\mu\nu}$
should vanish.

\section{Pion exchange contributions}
\label{sec:Pion}

In this Section we shall discuss the hadronic tensor and response
functions in the framework of the RFG model, {\it i.e.} for nucleons moving 
freely inside the nucleus with relativistic kinematics,
accounting for the effects introduced by pions in first-order perturbation
theory (one-pion-exchange).
We shall employ pseudovector (PV) coupling for the pion-nucleon
interactions. 

\subsection{Feynman diagrams and two-body currents}
\label{sec:Diag}

\begin{figure}[t]
\begin{center}
\leavevmode
\epsfbox[100 450 500 720]{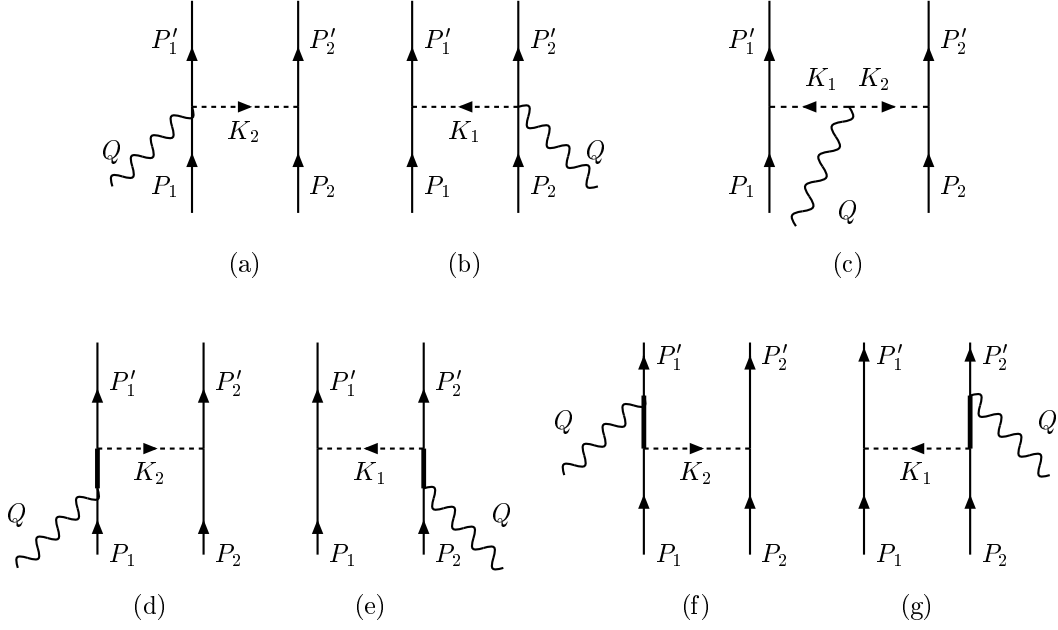}
\end{center} 
\caption{Free-space two-body current diagrams: Seagull (a,b), pion-in-flight
(c) and correlation (d-g) contributions.}
\end{figure}

We start by displaying in Fig.~1 the Feynman diagrams corresponding to
the free-space two-body currents with only one-pion-exchange. Using
the Feynman rules one writes down the corresponding expressions for
the two-particle current matrix elements.  Following standard
terminology, diagrams (a), (b) and (c) correspond to the usual
meson-exchange currents where the electromagnetic field is coupled
with the pionic current: in particular, diagrams (a) and (b) are
referred to as the seagull (contact) term and (c) as the
pion-in-flight term, respectively. Diagrams (d)-(g) contain an
intermediate virtual nucleon and give rise to the so-called
correlation currents.  In some work~\cite{Che71} it has been argued
that the contribution of these ``correlation'' diagrams is already
contained in the nuclear wave functions as solutions of the
Schr\"odinger equation.  However, within the RFG plus first-order
perturbation theory framework the unperturbed basis does not have any
such contributions and so within this approach they must be included
explicitly --- in fact, as we shall see, their inclusion is crucial
for fulfilling gauge invariance.

The relativistic expressions for the 
free-space two-particle currents are (isospin summations are understood)
\begin{itemize}
\item
Seagull or contact:
\end{itemize}
\ba
j_{\mu}^{S}(Q)&=& \left(\frac{f m}{V m_\pi}\right)^2 
             \frac{1}{\sqrt{E_{\np_1} E_{\np_2} E_{\np'_1} E_{\np'_2}}}
             i\epsilon_{3ab}
             \overline{u}(\np'_1)\tau_a\gamma_5\Kbar_1 u(\np_1)
             \frac{F_1^V}{K_1^2-m_\pi^2}
             \overline{u}(\np'_2)\tau_b\gamma_5\gamma_{\mu}u(\np_2)
\nonumber\\
             &+& (1 \leftrightarrow 2) \label{eq6}
\ea
\begin{itemize}
\item
Pion-in-flight:
\end{itemize}
\ba
j_{\mu}^{P}(Q) &=&  \left(\frac{f m}{V m_\pi}\right)^2 
             \frac{1}{\sqrt{E_{\np_1} E_{\np_2} E_{\np'_1} E_{\np'_2}}}
             i\epsilon_{3ab}
             \frac{F_\pi(K_1-K_2)_\mu}{(K_1^2-m_\pi^2)(K_2^2-m_\pi^2)}
\nonumber\\
     &\times&\overline{u}(\np'_1)\tau_a\gamma_5\Kbar_1 u(\np_1)
             \overline{u}(\np'_2)\tau_b\gamma_5\Kbar_2 u(\np_2)
\label{eq7}
\ea
\begin{itemize}
\item
Correlation:
\end{itemize}
\begin{eqnarray}
j_{\mu}^{C}(Q)
&=&           \left(\frac{f m}{V m_\pi}\right)^2 
              \frac{1}{\sqrt{E_{\np_1} E_{\np_2} E_{\np'_1} E_{\np'_2}}}
              \overline{u}(\np'_1)\tau_a\gamma_5\Kbar_1 u(\np_1)
              \frac{1}{K_1^2-m_\pi^2} \nonumber\\
& & \mbox{}\times \overline{u}(\np'_2)
    \left[    \tau_a\gamma_5\Kbar_1 
              S_F(P_2+Q)\Gamma^\mu(Q)
            + \Gamma^\mu(Q)S_F(P'_2-Q)
              \tau_a\gamma_5\Kbar_1
    \right]u(\np_2) \nonumber\\
& & \mbox{}+ (1\leftrightarrow2) \ . \label{eq8}
\end{eqnarray}
In the above, $V$ is the volume enclosing the system, $m_\pi$ is the
mass of the pion, $f$ the pion-nucleon coupling constant and
$E_{\np}=\sqrt{\np^2+m^2}$ the on-shell energy of a nucleon with
momentum $\np$; the four-momenta --- indicated by capital letters ---
are defined in Fig.~1 and $F_1^V$ and $F_\pi$ are the electromagnetic
isovector nucleon and pion form factors, respectively.  Furthermore,
$S_F(P)$ is the nucleon propagator and $\Gamma^\mu(Q)$ the
electromagnetic nucleon vertex, {\it i.e.,}
\begin{eqnarray}
S_F(P) &=& \frac{\Pbar + m}{P^2-m^2}
\label{NucleonVertex} \\
\Gamma^\mu(Q) &=&
F_1\gamma^\mu+\frac{i}{2m}F_2\sigma^{\mu\nu}Q_\nu \ ,\label{eq10}
\end{eqnarray}
$F_1$ and $F_2$ being the Dirac and Pauli form factors:
for these we use the Galster parameterization~\cite{Galster}.
Finally, the spinors (for brevity we denote $u(\np,s_p)$ by $u(\np)$) 
are normalized according to the Bjorken and Drell
convention~\cite{BD} and isospinors are not explicitly indicated.

The seagull and pion-in-flight currents shown above coincide with the
expressions given by Van Orden and Donnelly~\cite{VanOD81} if account
is taken for the different conventions used for the gamma matrix
$\gamma_5$ and for the metric.  Concerning the correlation current,
note that, at variance with~\cite{VanOD81}, it embodies both the
positive and negative energy components of the nucleon propagator.

An important point to be stressed is that the relativistic seagull,
pion-in-flight and correlation currents satisfy current conservation,
{\it i.e.} $Q_\mu J^\mu=0$, provided some assumptions are made for the
form factors involved in the various currents. To prove this statement
we start by evaluating the contraction of the four-momentum transfer
$Q^\mu$ with the correlation current $j_\mu^C$. It can be written as
\begin{equation}
Q^\mu j_\mu^C =  \left(\frac{f m}{V m_\pi}\right)^2 
             \frac{1}{\sqrt{E_{\np_1} E_{\np_2} E_{\np'_1} E_{\np'_2}}}
              \overline{u}(\np'_1)\tau_a\gamma_5\Kbar_1u(\np_1)
              \frac{1}{K_1^2-m_\pi^2} {\cal M}_a 
             +(1 \leftrightarrow 2)
\end{equation}
with ${\cal M}_a$ given by
\begin{equation}
{\cal M}_a =\overline{u}(\np'_2)
    \left[    \tau_a\gamma_5\Kbar_1 
              S_F(P_2+Q)\Qbar F_1
            + F_1 \Qbar S_F(P'_2-Q)
              \tau_a\gamma_5\Kbar_1
    \right]u(\np_2) \ ,
\end{equation}
where we have used the relation $Q^\mu \Gamma_\mu(Q) = F_1(Q)\Qbar$. 
After some algebra, involving the nucleon propagator and the Dirac spinors, 
${\cal M}_a$ can be further simplified leading to the form
\begin{equation}
{\cal M}_a=\overline{u}(\np'_2)
    \left[    \tau_a\gamma_5\Kbar_1 F_1 
            - F_1\tau_a\gamma_5\Kbar_1
    \right]u(\np_2) =\overline{u}(\np'_2)
    \left[\tau_a, F_1\right]
    \gamma_5\Kbar_1 
    u(\np_2) \ .
\end{equation}
In order to evaluate the commutator $[\tau_a,F_1]$
we decompose the nucleon form factor into its isoscalar and isovector pieces,
$F_1=\frac12\left( F_1^S+F_1^V\tau_3 \right)$. Then
\begin{equation}
\left[\tau_a,F_1\right] = -iF_1^V\epsilon_{3ab}\tau_b \ ,
\label{commutator}
\end{equation}
which entails the automatic conservation of the $\pi^0$ exchange current
($a$=3).
Using eq.~(\ref{commutator}) we can recast ${\cal M}_a$ as follows
\begin{equation}
{\cal M}_a = -i F_1^V \epsilon_{3ab}
\overline{u}(\np'_2)
    \tau_b\gamma_5\Kbar_1 
    u(\np_2) \ ;
\end{equation}
hence the divergence of the two-body correlation current matrix element
can finally be written as
\ba \label{div1}
Q^\mu j_\mu^C &=& 2m \left(\frac{f m}{V m_\pi}\right)^2 
             \frac{1}{\sqrt{E_{\np_1} E_{\np_2} E_{\np'_1} E_{\np'_2}}}
              i\epsilon_{3ab}
              \overline{u}(\np'_1)\tau_a\gamma_5 u(\np_1)
              \frac{F_1^V}{K_1^2-m_\pi^2}
\nonumber\\
&\times&              \overline{u}(\np'_2)
              \tau_b\gamma_5(\Qbar+2m) 
              u(\np_2)
             +(1 \leftrightarrow 2) \ .
\ea

The divergence of the seagull and pion-in-flight two-body current matrix 
elements can also be calculated in a straightforward way. The final result
reads
\ba \label{div2}
Q^\mu j_\mu^S &=&  -2m\left(\frac{f m}{V m_\pi}\right)^2 
             \frac{1}{\sqrt{E_{\np_1} E_{\np_2} E_{\np'_1} E_{\np'_2}}}
              i\epsilon_{3ab}
              \overline{u}(\np'_1)\tau_a\gamma_5 u(\np_1)
              \frac{F_1^V}{K_1^2-m_\pi^2}
\nonumber\\
&\times&
              \overline{u}(\np'_2)
              \tau_b\gamma_5\Qbar 
              u(\np_2)
             +(1 \leftrightarrow 2)
\ea

\ba \label{div3}
Q^\mu j_\mu^P &=& 4m^2 \left(\frac{f m}{V m_\pi}\right)^2 
             \frac{1}{\sqrt{E_{\np_1} E_{\np_2} E_{\np'_1} E_{\np'_2}}}
              i\epsilon_{3ab} F_\pi
              \frac{(K_1-K_2)\cdot Q}{(K_1^2-m_\pi^2)(K_2^2-m_\pi^2)}
\nonumber\\
&\times&      \overline{u}(\np'_1)\tau_a\gamma_5u(\np_1)
              \overline{u}(\np'_2)\tau_b\gamma_5u(\np_2) \ .
\ea
Then, by summing up the contributions given by the correlation 
(eq.~(\ref{div1})) and seagull (eq.~(\ref{div2})) 
currents and writing the four-momentum transfer
as $Q_\mu=(K_1+K_2)_\mu$, we finally obtain
\ba 
Q^\mu (j_\mu^C+ j_\mu^S) 
&=&  4m^2 \left(\frac{f m}{V m_\pi}\right)^2 
             \frac{1}{\sqrt{E_{\np_1} E_{\np_2} E_{\np'_1} E_{\np'_2}}} F_1^V
              i\epsilon_{3ab}
  \frac{(K_2-K_1)\cdot Q}{(K_1^2-m_\pi^2)(K_2^2-m_\pi^2)}
\nonumber\\
&\times&
            \overline{u}(\np'_1)\tau_a\gamma_5u(\np_1)
               \overline{u}(\np'_2)
              \tau_b\gamma_5 u(\np_2) \ ,
\ea
which cancels exactly the contribution of
pion-in-flight current in eq.~(\ref{div3}) provided the 
electromagnetic pion form
factor is chosen to be $F_\pi=F_1^V$, which we shall assume in this paper.
 
We have thus proven that the sum of the relativistic two-body matrix
elements of the MEC and of the correlation currents satisfies the
continuity equation if the same electromagnetic form factors enter in
all of the currents: namely one has $Q^\mu (j_\mu^C+ j_\mu^S +
j_\mu^P) = 0$.

\subsection{Particle-hole matrix elements}
\label{sec:phme}

In this paper we focus on the one-particle emission reactions within
the framework of the RFG. Therefore, one has to evaluate the matrix
element of the two-body current operators taken between the Fermi
sphere and a particle-hole state, namely
\begin{equation}
\langle ph^{-1}|{\hat j}_\mu|F\rangle = \sum_{s_k,t_k}\sum_{\nk \leq k_F}
           \left[\langle pk|{\hat j}_\mu|h k\rangle -
                 \langle pk|{\hat j}_\mu|kh\rangle \right],
\label{eq22}
\end{equation}
where the sum $\sum_{\nk \leq k_F}$ in the thermodynamic limit becomes
an integral over the momentum in the range $0\leq k\leq k_F$, with
$k_F$ the Fermi momentum, and over the angular variables
$\theta_k,\phi_k$.  The first and second terms in eq.~(\ref{eq22})
represent the direct and exchange contribution to the matrix element,
respectively.  It is well-known that the direct term vanishes for the
MEC (seagull and pion-in-flight) currents (see~\cite{Ama98}) because
of both spin-isospin traces (in spin-isospin saturated systems) and
the presence of a pion carrying zero momentum.  It is easily verified
that, for the same reasons, the direct matrix element of the
correlation current also does not contribute to the responses.

\begin{figure}[t]
\begin{center}
\leavevmode
\epsfbox[100 450 500 720]{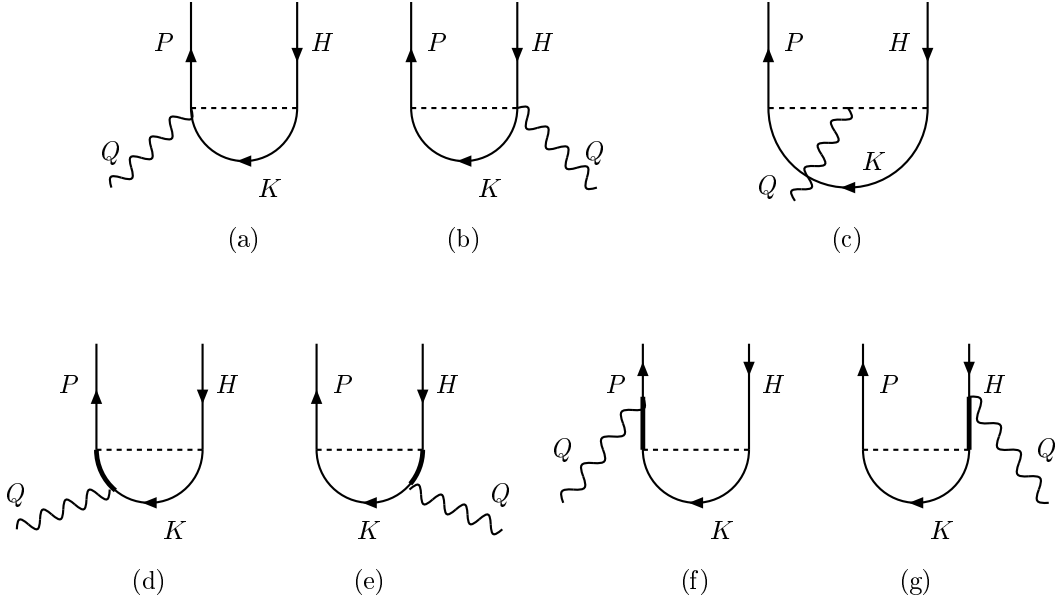}
\end{center}
\caption{Feynman diagrams representing the seagull (a and b), pion-in-flight
(c), vertex (d and e) and self-energy of the hole (f) and particle (g)
particle-hole matrix elements.}
\end{figure}

Hence only the exchange term remains in the particle-hole matrix
elements for the two-body current operators. The corresponding Feynman
diagrams are displayed in Fig.~2. Diagrams (a-b) and (c) correspond to
the seagull (contact) and pion-in-flight terms, respectively.
Diagrams (d-e) represent instead the correlation contributions.  Here
we distinguish the self-energy insertions on the nucleonic lines (f,g)
and the exchange of a pion between a particle and a hole line (d,e).

Starting from the general expressions in
eqs.~(\ref{eq6},\ref{eq7},\ref{eq8}) we can now evaluate the
particle-hole matrix elements.  Carrying out explicitly the sums over
the spin, $s_k$, and isospin, $t_k$, variables and after some
laborious, but straightforward, algebra we finally get
\begin{itemize}
\item
Seagull
\end{itemize}
\ba
\langle ph^{-1}|{\hat j}_{\mu}^{S}|F\rangle
&=& -\frac{m f^2}{V^2 m_\pi^2\sqrt{E_{\np} E_{\nh}}} F_1^{(V)}
i\varepsilon_{3ab}
\nonumber\\
&& \kern -2cm 
\mbox{}\times
 \sum_{\nk}\frac{m}{E_{\nk}}
              \overline{u}(\np)\tau_a\tau_b
\left\{ \frac{(\Kbar-m)\gamma_\mu}{(P-K)^2-m_\pi^2}
      + \frac{\gamma_\mu (\Kbar-m)}{(K-H)^2-m_\pi^2} 
\right\}u(\nh)
\label{Sph}
\ea

\begin{itemize}
\item
Pion-in-flight
\end{itemize}
\ba
\langle ph^{-1}|{\hat j}_{\mu}^{P}|F\rangle
&=&
2m \frac{m f^2}{V^2 m_\pi^2\sqrt{E_{\np} E_{\nh}}} F_1^{(V)}
i\varepsilon_{3ab} 
\nonumber\\
&&
\kern -2cm
\mbox{}\times
\sum_{\nk}
        \frac{m}{E_{\nk}}\frac{(Q+2H-2K)_\mu}{[(P-K)^2-m_\pi^2]
         [(K-H)^2-m_\pi^2]}
\overline{u}(\np)\tau_a(\Kbar-m)
              \tau_b u(\nh)
\label{Pph}
\ea

\begin{itemize}
\item
Correlation
\end{itemize}
\ba
& & \langle ph^{-1}|{\hat j}_\mu^{C}|F\rangle
=           -\frac{1}{2m} \frac{m f^2}{V^2 m_\pi^2 \sqrt{E_{\np} E_{\nh}}}
             \sum_{\nk}\frac{m}{E_{\nk}}
              \overline{u}(\np) \left\{ \tau_a\gamma_5
              \frac{\Pbar-\Kbar}{(P-K)^2-m_\pi^2} (\Kbar + m) 
              \right.
\nonumber\\
& & \times 
    \left[    \tau_a\gamma_5(\Pbar-\Kbar)
              S_F(H+Q)\Gamma_\mu(Q)
            + \Gamma_\mu(Q)S_F(K-Q)
              \tau_a\gamma_5
              (\Pbar-\Kbar)
    \right] \nonumber\\
& &            + \left[    \tau_a\gamma_5(\Kbar-\Hslash)
              S_F(K+Q)\Gamma_\mu(Q)
            + \Gamma_\mu(Q)S_F(P-Q)
              \tau_a\gamma_5(\Kbar-\Hslash)
    \right] 
\nonumber\\
& & \times \left.
              (\Kbar + m)\tau_a\gamma_5
              \frac{\Kbar-\Hslash}{(K-H)^2-m_\pi^2}
  \right\} u(\nh)
\ \label{corr1}.
\ea
The first and last terms in eq.~(\ref{corr1}) represent the so-called 
self-energy (s.e.) matrix elements (diagrams (g) and (f) in Fig.~2), whereas
the second and third terms correspond to the vertex correlations
(v.c., sometimes referred to as ``exchange'') matrix elements 
(diagrams (e) and (d)). We examine them separately, writing accordingly 

\be
\langle ph^{-1}|j_\mu^C|F\rangle =
\langle ph^{-1}|j_\mu^C|F\rangle_{s.e.}+
\langle ph^{-1}|j_\mu^C|F\rangle_{v.c.} \ .
\ee

The self-energy ph matrix element can be split into two parts, ${\cal
H}^\mu_p$ and ${\cal H}^\mu_h$.  The former corresponds to the diagram
with the pion inserted in the particle line (Fig.~2g), whereas the
latter describes the diagram with the pion inserted in the hole line
(Fig.~2f). They are

\be
{\cal H}^\mu_p =   -\frac{3 m f^2}{V^2 m_\pi^2\sqrt{E_{\np} E_{\nh}}} 
\sum_{\nk}\frac{m}{E_{\nk}}
\overline{u}(\np)
 (\Kbar- m)\frac{\Pbar-\Kbar}{(P-K)^2-m_\pi^2}
              S_F(P)\Gamma^\mu(Q) u(\nh)
\label{Sp}
\ee
\be
{\cal H}^\mu_h =   \frac{3 m f^2}{V^2 m_\pi^2\sqrt{E_{\np} E_{\nh}}} 
\sum_{\nk}\frac{m}{E_{\nk}}
\overline{u}(\np)
\Gamma^\mu(Q)S_F(H)\frac{(\Kbar-\Hslash)}{(K-H)^2-m_\pi^2}(\Kbar-m)u(\nh)\ ,
\label{Sh}
\ee
where the isospin trace has been performed.

Similarly, the vertex ph matrix element splits into two terms ${\cal
F}^\mu$ and ${\cal B}^\mu$ representing the forward- and
backward-going contributions, respectively (Fig.~2d and 2e). They are

\be
{\cal F}^\mu
=           -\frac{mf^2}{V^2 m_\pi^2\sqrt{E_{\np} E_{\nh}}} 
             \sum_{\nk}\frac{m}{E_{\nk}} \overline{u}(\np)
              \gamma_5(\Kbar-\Hslash)
              S_F(K+Q)\tau_a\Gamma^\mu(Q)\tau_a \gamma_5 
              \frac{(\Kbar - m)}{(K-H)^2-m_\pi^2} u(\nh)
\label{F}
\ee
\be
{\cal B}^\mu =
          -\frac{mf^2}{V^2 m_\pi^2\sqrt{E_{\np} E_{\nh}}} 
           \sum_{\nk}\frac{m}{E_{\nk}} \overline{u}(\np)
              \frac{\Kbar-m}{(P-K)^2-m_\pi^2} \gamma_5 
             \tau_a\Gamma^\mu(Q)\tau_a S_F(K-Q)
              \gamma_5
              (\Pbar-\Kbar) u(\nh)\ .
\label{B}
\ee

Finally, splitting also the electromagnetic nucleon operator $\Gamma^\mu$ 
into its isoscalar and isovector parts, one obtains the isoscalar 
and isovector
contributions to the self-energy and vertex ph matrix elements.
The final results can be cast in the form
\ba
{\cal H}^{\mu(S,V)}_p 
&=&            -\frac{3f^2}{2 V^2 m_\pi^2 \sqrt{E_{\np} E_{\nh}}} 
\nonumber\\
&\times&              \sum_{\nk}\frac{m}{E_{\nk}}
 \overline{u}(\np)
 \frac{\Pbar-\Kbar}{(P-K)^2-m_\pi^2} (\Kbar - m) (\Pbar-\Kbar)
              S_F(P)\Gamma^{\mu(S,V)}(Q) u(\nh)\nonumber \\
\label{se1} 
\ea
and
\ba
{\cal H}^{\mu(S,V)}_h 
&=&            -\frac{3f^2}{2 V^2 m_\pi^2 \sqrt{E_{\np} E_{\nh}}} 
\nonumber\\
&\times&              \sum_{\nk}\frac{m}{E_{\nk}}
\overline{u}(\np)
\Gamma^{\mu(S,V)}(Q)S_F(H)(\Kbar-\Hslash)(\Kbar-m)
\frac{(\Kbar-\Hslash)}{(K-H)^2-m_\pi^2} u(\nh), \nonumber \\
 \label{se2}
\ea
for the self-energy matrix elements and
\ba
{\cal F}^{\mu(S)}
&=&            -\frac{3 m f^2}{V^2 m_\pi^2 \sqrt{E_{\np} E_{\nh}}} ç
\nonumber\\
&\times&               \sum_{\nk}\frac{m}{E_{\nk}}
 \overline{u}(\np)
 \gamma_5(\Kbar-\Hslash)
              S_F(K+Q)\Gamma^{\mu(S)}(Q) \gamma_5
              \frac{(\Kbar - m)}{(K-H)^2-m_\pi^2} u(\nh)
\\
{\cal B}^{\mu(S)}
&=&            -\frac{3 m f^2}{V^2 m_\pi^2 \sqrt{E_{\np} E_{\nh}}} 
\nonumber\\
&\times& \sum_{\nk}\frac{m}{E_{\nk}}
\overline{u}(\np)
              \frac{\Kbar-m}{(P-K)^2-m_\pi^2} 
             \gamma_5 \Gamma^{\mu(S)}(Q)S_F(K-Q) \gamma_5
              (\Pbar-\Kbar)u(\nh) \ ,
\ea
for the isoscalar and
\ba
& &{\cal F}^{\mu(V)}
=            -\frac{m f^2}{V^2m_\pi^2 \sqrt{E_{\np} E_{\nh}}}\nonumber \\
&\times &
 \sum_{\nk}\frac{m}{E_{\nk}}
 \overline{u}(\np)
 \gamma_5(\Kbar-\Hslash)
              S_F(K+Q)\Gamma^{\mu(V)}(Q)(\tau_3+i\varepsilon_{3ab}
              \tau_a\tau_b)
              \gamma_5\frac{(\Kbar - m)}{(K-H)^2-m_\pi^2} u(\nh)
\nonumber \\
& & \label{FV} \\
& &{\cal B}^{\mu(V)}
=            -\frac{m f^2}{V^2m_\pi^2 \sqrt{E_{\np} E_{\nh}}} 
\nonumber \\
&\times & \sum_{\nk}
\frac{m}{E_{\nk}}
\overline{u}(\np)
              \frac{\Kbar-m}{(P-K)^2-m_\pi^2} 
            \gamma_5 \Gamma^{\mu(V)}(Q)(\tau_3+i\varepsilon_{3ab}
            \tau_a\tau_b)
 S_F(K-Q) \gamma_5 (\Pbar-\Kbar)u(\nh) \ , \nonumber \\
& & \label{BV}
\ea
for the isovector vertex ph matrix elements. Interestingly,
as far as isospin traces are concerned,
in the vertex matrix element
the isoscalar/isovector ratio is -3, whereas in the self-energy case
it is 1
\footnote{The latter result stems from the relation 
$\tau_3+i\varepsilon_{3ab}\tau_a\tau_b = -\tau_3$; however we prefer to
leave the isospin structure of the isovector exchange as in 
(\ref{FV},\ref{BV}) since it makes more transaparent the 
self-energy--exchange
cancellation in the continuity equation, as shown in appendix A.}.
Note that the MEC (pion-in-flight and seagull) ph matrix elements are purely
isovector, whereas the vertex and self-energy correlations get both isoscalar 
and isovector contributions.

Note that the v.c. and s.e. ph matrix elements involve
the nucleon propagator $S_F(P)$ which in some situations may imply 
the occurrence of singularities. In the case of the vertex
diagrams, the four-momenta appearing in the propagators are
$K+Q$ and $K-Q$ for the forward- (Fig.~2d) and backward-going (Fig.~2e)
contributions, respectively, and an integration over $\nk$ should be
done. For $q\geq 2 k_F$ (no Pauli blocking) it can be proven 
(see appendix B) 
that only the forward diagram contains a pole, {\it i.e.,} a value of the
inner momentum $\nk$ exists such that the nucleon
carrying a four-momentum $K+Q$ is on-shell. 
In this situation the forward vertex ph matrix element is 
evaluated by taking 
the principal value in the integral over $\cos\theta_k$. In the case of the
backward-going diagram the nucleon propagator $S_F(K-Q)$ has no
singularity for the kinematics in which we are interested. 

The case of the self-energy diagrams is different. Here
the ph matrix element diverges, as the nucleons
described by the Feynman propagators $S_F(P)$ and $S_F(H)$ are forced to
be on-shell by energy and momentum conservation. Accordingly this term 
should in principle be renormalized. In~\cite{Ama01} we present a
detailed analysis of the different aspects of the impact of the
self-energy term on the nuclear responses within the relativistic framework. 
A comparison with the non-relativistic reduction is also carried out there.
Here we confine ourselves to computing the relativistic response
functions through the hadronic tensor $W^{\mu\nu}$. As stated in 
Section~\ref{sec:Form},
$W^{\mu\nu}$ can be evaluated using the current matrix elements or the 
polarization propagator. In the first instance, however,
the presence of divergent ph matrix elements may lead to wrong responses or 
inconsistencies in the theory.
In this paper, to avoid this happening, we calculate the s.e. contribution 
to the response with the polarization propagator. 

An important point here is the fact that the hadronic 
tensor thus obtained, partly through the 
ph matrix elements and partly through the polarization propagator, is gauge
invariant. This may be somewhat surprising because we show in appendix 
A that current conservation is already obtained at the level of the 
MEC and correlation ph matrix elements: hence the one-body current 
ph matrix element also has to be conserved. 
However, this occurs only in zeroth order of perturbation theory. To be 
dealt with properly, clearly the situation requires the 
renormalization of the ph energies and Dirac spinors.
Only then does it become possible to set up a renormalized s.e. current
which leads to a hadronic tensor that coincides with the one obtained here 
through the polarization propagator~\cite{Ama01}.

\section{Nuclear hadronic tensor and response functions}
\label{sec:tensor}

In this Section we analyze the response functions 
for one-particle emission reactions within the RFG model.
As discussed in the previous Section, the ph matrix elements 
corresponding to the different pionic diagrams are all well-defined 
except for the self-energy term which diverges. Therefore, 
on the one hand, in the 
case of the one-body, MEC and vertex correlation diagrams
we evaluate the hadronic tensor starting from the current ph matrix 
elements (Fig.~2). On the other hand, for the self-energy 
diagrams we calculate the polarization propagator $\Pi^{\mu\nu}$. 

The analysis of the nuclear hadronic tensor set up with 
the ph matrix elements
has been presented in detail, within the RFG model, in previous
papers~\cite{ADM90,Barb93}. Hence, here we simply summarize 
the results needed for later discussions. The hadronic tensor that arises from 
the interference of the single-nucleon  current, $j_{s.n.}^{\mu}$  with the 
one-pion-exchange current $j_a^{\mu}$, with $a=S$ (seagull),
$P$ (pion-in-flight) and $v.c.$ (vertex), is for the RFG model with $Z=N$
\be
W^{\mu\nu}=\frac{3Z}{8\pi k_F^3q}
\int_{h_0}^{k_F}
h dh (\omega+E_{\nh}) 
\int_0^{2\pi}d\phi_h
\sum_{s_p,s_h}
2{\rm Re}\,
\left[\langle ph^{-1}|{\hat j}^\mu_{s.n.}|F\rangle^\ast
\langle ph^{-1}|{\hat j}^\nu_a|F\rangle \right] \ , \label{eq38}
\ee
where $\langle ph^{-1}|{\hat j}^\mu_{s.n.}|F\rangle=\frac{m}{\sqrt{E_{\np} E_{\nh}}}
\overline{u}({\bf p})
\Gamma^\mu u({\bf h})$ is the single-nucleon
ph matrix element with $\Gamma^\mu$ the electromagnetic nucleon 
current from eq.~(\ref{eq10}) and $\langle ph^{-1}|\hat j^\nu_a|F\rangle$ is the
ph matrix element for the seagull, pion-in-flight or vertex
current as given in eqs.~(\ref{Sph}), (\ref{Pph}) and (\ref{F}-\ref{B}), 
respectively.

Note that in eq.~(\ref{eq38}) the integral  over 
hole polar angle,
$\cos\theta_h$, has been performed explicitly by exploiting
the energy conserving $\delta$-function. This fixes the minimum momentum of
the hole according to
\be
h_0=m\sqrt{\varepsilon_0^2-1}
\ ,
\kern 4em
\varepsilon_0=
{\rm Max}\,
\left\{ \varepsilon_F-2\lambda,
\kappa\sqrt{1+\frac{1}{\tau}}-\lambda
\right\} \ ,
\label{neededforpsi}
\ee
where the usual dimensionless variables
\be
\lambda=\frac{\omega}{2m},\,\,\,\, \tau=\frac{|Q^2|}{4m^2},\,\,\,\,
\kappa=\frac{q}{2m},\,\,\,\,\varepsilon_F=\frac{E_F}{m},\,\,\,\,
\eta=\frac{h}{m}
\ee
have been introduced and 
$E_F=\sqrt{k_F^2+m^2}$ is the Fermi energy. Moreover, the hole
three-momentum 
\be
{\bf h}=h\left(\sin\theta_0\cos\phi_h,\sin\theta_0\sin\phi_h,
\cos\theta_0\right)\ ,
\ee
involved in the hadronic tensor, must be evaluated for the following
specific value of the polar angle
\be
\cos\theta_0=\frac{\lambda\varepsilon-\tau}{\eta\kappa} \ .
\ee

The hadronic tensor, as was the case for the current, can be also split into 
isoscalar and isovector parts, since there is no interference between
the two isospin channels.

An important issue relates to the form factor
of the $\pi NN$ vertex, $\Gamma_\pi$, which incorporates some 
aspects of the short-range
physics affecting the pionic correlation. In all of the above
expressions $\Gamma_\pi$ has not been explicitly indicated 
for sake of simplicity. Here we recall that its inclusion in the 
ph current matrix 
elements is not without consequences in connection with gauge invariance. 
In fact, in this case, the results quoted in appendix A are no longer
valid unless new terms are added to the MEC
(see~\cite{MECpapers,Barb93,Barb94}). Lacking a fundamental 
theory for $\Gamma_\pi$, in the calculations reported in the next 
Section we use the phenomenological expression
\be
\Gamma_\pi(P)=\frac{\Lambda^2-m_\pi^2}{\Lambda^2-P^2}
\label{eq44}
\ee
with $\Lambda=1.3$ GeV. As long as the dependence upon
$\Lambda$ is not too strong the gauge invariance of the theory should not be
too badly affected. Within a non-relativistic approach for the pion currents,
a detailed discussion on the breakdown of the gauge invariance due to
eq.~(\ref{eq44}), and on the dependence of the responses upon the cutoff
value can be found in~\cite{Barb93,Barb94}.

Now, as already discussed, a crucial point to be emphasized is that
the self-energy ph matrix element is divergent. Hence it should be 
renormalized
to be used in the evaluation of the hadronic tensor. Referring the
reader to~\cite{Ama01} for a thorough discussion on this subject,
here we show how the hadronic tensor is 
built by evaluating the polarization propagator corresponding 
to the two self-energy diagrams shown in Fig.~3.

\begin{figure}[t]
\begin{center}
\leavevmode
\epsfbox[100 550 500 720]{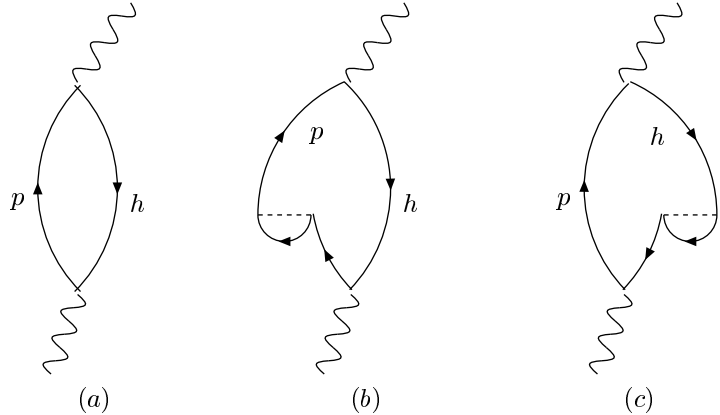}
\end{center}
\caption{
Feynman diagrams of the free (a) and first-order 
self-energy (b and c) polarization  propagator.}
\end{figure}

  From the Feynman diagrams with the self-energy $\Sigma$
inserted in the particle
or in the hole lines (Fig.~3), the polarization propagator is 
\begin{eqnarray}
\Pi^{\mu\nu}(Q) 
&=& -i\int\frac{dh_0 d^3h}{(2\pi)^4} 
     {\rm Tr} \left\{
     \Gamma^{\mu}(Q) S(H)\Sigma(H) S(H) \Gamma^{\nu}(-Q) S(H+Q) \right.
\nonumber\\ 
&& \left. \kern 2cm \mbox{} 
     +\Gamma^{\mu}(Q) S(H) \Gamma^{\nu}(-Q)S(H+Q)\Sigma(H+Q)S(H+Q)
   \right\} \ .
\end{eqnarray}
In the above the nucleon propagator within the RFG framework reads

\ba
S(K)& =& \frac{\Kbar+m}{K²-m²+i\epsilon}
      + 2\pi i (\Kbar+m)
        \theta(k_F-k)
        \delta(K^2-m^2)\theta(k_0) \nonumber \\
&=& (\Kbar+m)
      \left[ \frac{\theta(k-k_F)}{K^2-m^2+i\epsilon}
            +\frac{\theta(k_F-k)}{K^2-m^2-i\epsilon k_0}
      \right] \ .\label{eq45}
\ea
Concerning the self-energy one has
\be
\Sigma(P)=-\frac{f^2}{Vm_\pi^2}\sum_{\nk,s_k,t_k}\frac{m}{E_{\nk}}
\tau_a\gamma_5(\Pbar-\Kbar)\frac{u(\nk)\overline{u}(\nk)}{(P-K)^2-m_\pi^2}
\tau_a\gamma_5(\Pbar-\Kbar)  \ .
\ee

In the following we illustrate the calculation by treating in detail 
the case of the self-energy in the hole line. The analysis of the 
diagram with the s.e. insertion
on the particle line is similar and then we shall just quote the final
result. In appendix C we discuss the more general case with any number
of self-energy insertions in the particle and/or in the hole lines.

Using the nucleon propagator in eq.~(\ref{eq45}), we can write the polarization 
propagator with the s.e. on the hole line as follows
\begin{eqnarray}
\Pi^{\mu\nu}(Q)
&=& -i {\rm Tr} \int\frac{dh_0 d^3h}{(2\pi)^4}  
    \Gamma^{\mu}(Q)(\Hslash+m)\Sigma(H) (\Hslash+m)
\nonumber\\
&\times&      \left[ \frac{\theta(h-k_F)}{(H^2-m^2+i\epsilon)^2}
            +\frac{\theta(k_F-h)}{(H^2-m^2-i\epsilon h_0)^2}
      \right]    \Gamma^{\nu}(-Q)(\Hslash+\Qbar+m) 
\nonumber\\
&& \times
      \left[ \frac{\theta(|\nh+\nq|-k_F)}{(H+Q)^2-m^2+i\epsilon}
            +\frac{\theta(k_F-|\nh+\nq|)}{(H+Q)^2-m^2-i\epsilon (h_0+q_0)}
      \right].
\end{eqnarray}
Then, exploiting the identity~\cite{ADM90}
\begin{equation}
\frac{1}{(H^2-m^2-i\epsilon)^2}
= \left.\frac{d}{d\alpha}\right|_{\alpha=0}
  \frac{1}{H^2-\alpha-m^2-i\epsilon},
\end{equation}
and subtracting the (infinite) vacuum polarization (see appendix C) we get
\begin{eqnarray}
\Pi^{\mu\nu}(Q)
&=& -i\left.\frac{d}{d\alpha}\right|_{\alpha=0}
     {\rm Tr} \int\frac{dh_0 d^3h}{(2\pi)^4}  
    \Gamma^{\mu}(Q)(\Hslash+m)\Sigma(H) (\Hslash+m)
    \Gamma^{\nu}(-Q)(\Hslash+\Qbar+m) 
\nonumber\\
&&   \times
     \left[\frac{\theta(k_F-h)}{H^2-\alpha-m^2-i\epsilon h_0}\right]
     \left[
     \frac{\theta(|\nh+\nq|-k_F)}{(H+Q)^2-m^2+i\epsilon}\right] \ .
\end{eqnarray}
 
Now the integral over the hole energy $h_0$ can be computed.
For this purpose we rewrite the denominators of the hole and particle
propagators in the form
\ba
H^2-\alpha-m^2-i\epsilon h_0 &=& H^2-m'{}^2-i\epsilon h_0 =
(h_0+E'_\nh-i\epsilon )(h_0-E'_\nh-i\epsilon) \\
(H+Q)^2-m^2+i\epsilon &=& 
(h_0+q_0+E_{\np}-i\epsilon)(h_0+q_0-E_{\np}+i\epsilon) \ ,
\ea
where we have introduced a mass

\begin{equation}
m' = m'(\alpha)  \equiv \sqrt{\alpha+m^2} \\
\end{equation}
and an energy

\begin{equation}
E'_\nh = E'_\nh(\alpha) \equiv \sqrt{\nh^2+m'{}^2}
\end{equation}
that depend upon the parameter $\alpha$ and we have used the relations
$\np = \nh+\nq$, $E_{\np} = \sqrt{\np^2+m^2}$. 
Therefore the poles are located at the points

\begin{eqnarray}
h_0 &=& \pm E'_\nh+i\epsilon\\
h_0 &=& \pm(E_\np-i\epsilon)-q_0 
\end{eqnarray}
and the integral over $h_0$ can be computed by closing the integration
path in the lower
half-plane, so that only the pole at $h_0=E_\np-i\epsilon-q_0$ contributes.
Then the polarization propagator is found to be

\begin{eqnarray}
& &\Pi^{\mu\nu}(Q) \nonumber \\
&=& -\left.\frac{d}{d\alpha}\right|_{\alpha=0}
     {\rm Tr} \int\frac{d^3h}{(2\pi)^3} 
    \Gamma^{\mu}(Q)(\Hslash+m)\Sigma(H) (\Hslash+m)
    \Gamma^{\nu}(-Q)\left. (\Hslash+\Qbar+m)\right|_{h_0=E_\np-q_0} 
\nonumber\\
&&   \kern 2cm \times
     \frac{1}{2E_\np}
     \frac{\theta(k_F-h)\theta(|\nh+\nq|-k_F)}
          {(E_\np-q_0)^2-\nh^2-m'{}^2-i\epsilon(E_\np-q_0)} \ .
\end{eqnarray}
The hadronic tensor then follows from formula~(\ref{eq4}) and reads

\begin{eqnarray}
&&W^{\mu\nu}=
-\frac{1}{\pi}
{\rm Im}\, \left[\Pi^{\mu\nu}(Q)\right] \nonumber \\
&&=\frac{3\pi^2 Z}{k_F^3} \left.\frac{d}{d\alpha}\right|_{\alpha=0}
     {\rm Tr} \int\frac{d^3h}{(2\pi)^3} 
    \Gamma^{\mu}(Q)(\Hslash+m)\Sigma(H) (\Hslash+m)
    \Gamma^{\nu}(-Q)\left. (\Hslash+\Qbar+m)\right|_{h_0=E_\np-q_0} 
\nonumber\\
&&   \times
     \frac{1}{2E_\np}
     \theta(k_F-h)\theta(|\nh+\nq|-k_F)
     \frac{1}{2(E_\np-q_0)}
     \delta(E_\np-q_0-E'_\nh) \ ,
\end{eqnarray}
where we have exploited the inequality $E_\np-q_0 >0$  and
the relation $\delta(x^2-x_0^2)=\delta(x-x_0)/2x_0$ for $x_0>0$.

Finally, using the energy-conserving delta-function and setting
$p_0=E_{\np}$, we obtain
\begin{eqnarray}
W^{\mu\nu}
&=& \left.\frac{3\pi^2 Z}{k_F^3} \frac{d}{d\alpha}\right|_{\alpha=0}
     {\rm Tr} \int\frac{d^3h}{(2\pi)^3}  
    \Gamma^{\mu}(Q)(\Hslash+m)\Sigma(H) (\Hslash+m)
    \Gamma^{\nu}(-Q)\left. (\Pbar+m)\right|_{h_0=E'_\nh(\alpha)}
\nonumber\\
&&   \kern 2cm \times
     \frac{1}{2E_\np}
     \theta(k_F-h)\theta(|\nh+\nq|-k_F)
     \frac{1}{2E'_\nh(\alpha)}
     \delta(E_\np-q_0-E'_\nh(\alpha)) \ ,
\end{eqnarray}
an expression that coincides with the one given by eq.~(\ref{eq106}) in
appendix C. 

Using a similar analysis, the hadronic tensor with
the self-energy inserted in the particle line turns out to read
(see appendix C)
\begin{eqnarray}
W^{\mu\nu}
&=& \left.\frac{3\pi^2 Z}{k_F^3} \frac{d}{d\beta}\right|_{\beta=0}
     {\rm Tr} \int\frac{d^3h}{(2\pi)^3}  
    \Gamma^{\mu}(Q)(\Hslash+m)\Gamma^{\nu}(-Q)(\Pbar+m)
    \Sigma(P)\left. (\Pbar+m)\right|_{p_0=E'_\np(\beta)}
\nonumber\\
&&   \kern 2cm \times
     \frac{1}{2E'_\np(\beta)}
     \theta(k_F-h)\theta(|\nh+\nq|-k_F)
     \frac{1}{2E_\nh}
     \delta(E'_\np(\beta)-q_0-E_\nh)
\end{eqnarray}
with $E'_\np(\beta)=\sqrt{\np^2+m'^2}$ and $m'=m'(\beta)=\sqrt{\beta+m^2}$.

In the above expressions, after the derivatives with respect to the 
parameters $\alpha$ and $\beta$ are taken,
the integral over the hole 
polar angle $\cos\theta_h$ can be performed analytically by exploiting the 
$\delta$-function. Hence the s.e. contribution to
the hadronic tensor can finally be expressed as a double integral.
Since the self-energy $\Sigma$ involves a triple integral, the contribution
to hadronic tensor turns out to be a 5-dimensional integral, to be 
carried out numerically.

\section{Results}
\label{sec:res}

In this Section we report the numerical results obtained for
the pionic MEC (pion-in-flight and seagull)
and for the correlation (vertex and self-energy)
contributions to the quasielastic peak in the 1p-1h sector.
The calculation is fully relativistic. We have taken $Z=N=20$ and set
$k_F=237$ MeV/c, which is representative of nuclei in the vicinity of
$^{40}$Ca.

The 5-dimensional integrations of the MEC and correlation responses
implicit in eq.~(\ref{eq38}) have been performed numerically. 
The reliability of the numerical procedure has been proven by checking that 
the free RFG responses coincide with their analytical expressions 
(see, e.g., \cite{Don92}).
Moreover the transverse MEC responses have been compared with the 
non-relativistic calculation developed in
\cite{Ama94b}, where the seagull ph matrix element is evaluated
analytically, while the pion-in-flight contribution is reduced to a
one-dimensional integral: both calculations give the same results for
$q$ and $k_F$ small.

\begin{figure}[p]
\begin{center}
\leavevmode
\epsfbox[100 495 500 685]{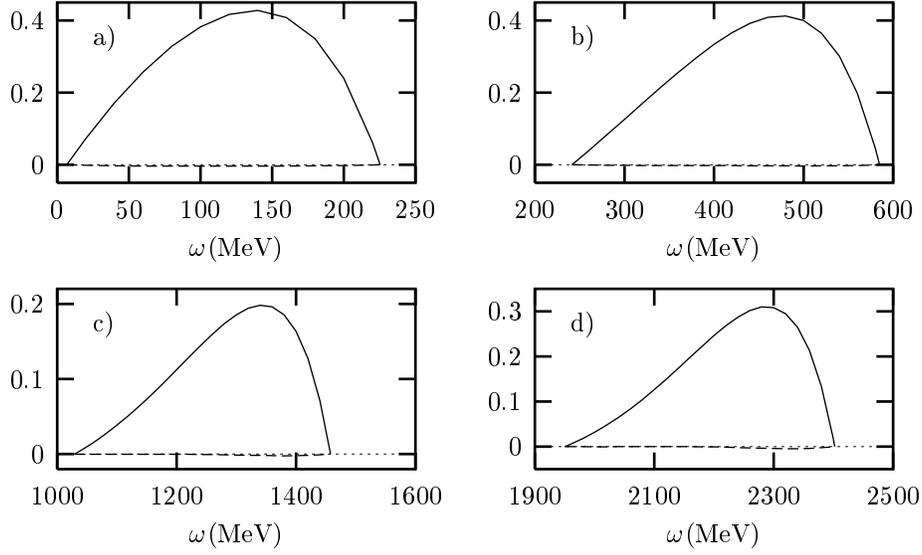}
\end{center}
\caption{
Longitudinal response versus $\omega$. Solid: free; dashed: MEC contribution.
Here and in all of the figures to follow the nucleus is $^{40}$Ca 
with $k_F$=237 MeV/c 
and the units are 
$10^{-1}$ MeV$^{-1}$ at $q$=0.5 GeV/c (panel a),
$10^{-2}$ MeV$^{-1}$ at $q$=1 GeV/c (panel b), 
$10^{-3}$ MeV$^{-1}$ at $q$=2 GeV/c (panel c) and 
$10^{-4}$ MeV$^{-1}$ at $q$=3 GeV/c (panel d).
}
\end{figure}

\begin{figure}[p]
\begin{center}
\leavevmode
\epsfbox[100 495 500 685]{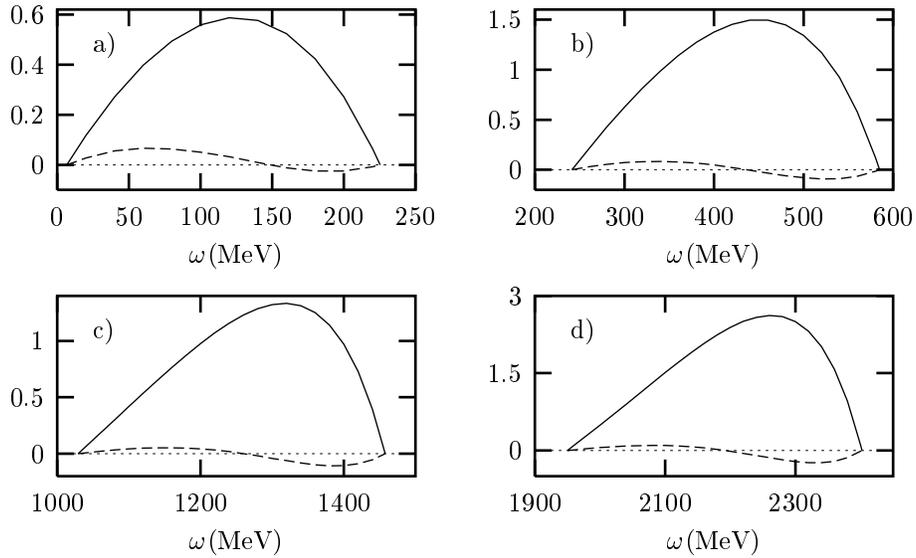}
\end{center}
\caption{Transverse response versus $\omega$ at $q$=0.5 (a),
1 (b), 2 (c) and 3 (d) GeV/c. Solid: free; dashed: 
MEC contribution. 
}
\end{figure}

\subsection{MEC}
\label{sec:MEC}

With regard to the MEC we have found the following:

\begin{itemize}

\item
They are almost irrelevant in the longitudinal channel, 
whereas they are small but not insignificant in the transverse one.
This outcome emerges from inspection of
Figs.~4 and 5, where the MEC contribution (dashed curves) to the L and
T responses is displayed for various values of $q$ together with the free
responses (solid). While in the L channel the MEC are hardly
visible, in the T channel they contribute somewhat more, although the
contributions there typically only amount to about 5--10\%, 
depending upon $q$ and $\omega$
(see later). This dominance of the spin interaction between the
photon and the MEC prompts us to concentrate on the 
transverse channel.

\begin{figure}[ht]
\begin{center}
\leavevmode
\epsfbox[100 495 500 690]{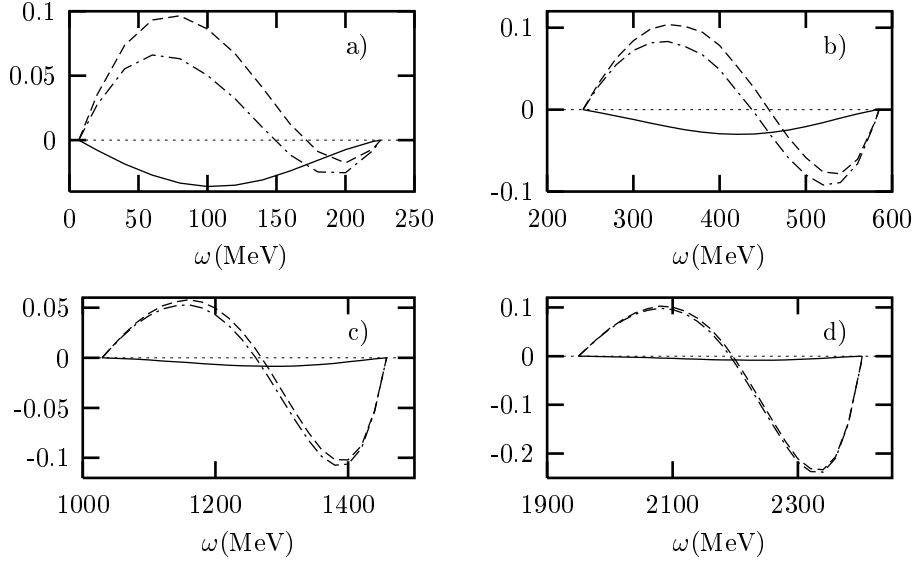}
\end{center}
\caption{
MEC contribution to transverse response versus $\omega$ at $q$=0.5 (a),
1 (b), 2 (c) and 3 (d) GeV/c. 
Solid: pion-in-flight; dashed: seagull; dot-dashed: total MEC.
}
\end{figure}

\begin{figure}[hp]
\begin{center}
\leavevmode
\epsfbox[100 550 500 690]{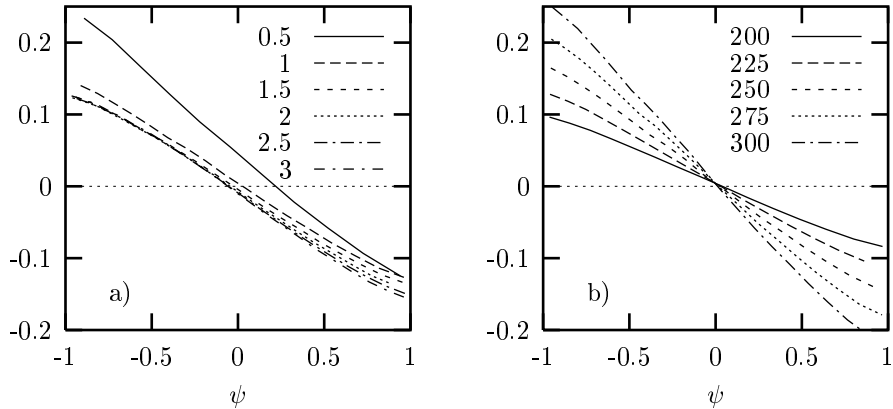}
\end{center}
\caption{
The ratio MEC/free in the transverse channel plotted versus 
$\psi$ at $k_F$=237 
MeV/c for various values of $q$ (in GeV/c) in panel a and
for various values of $k_F$  (in MeV/c) at $q$=1 GeV/c in panel b.
}
\end{figure}

\begin{figure}[hp]
\begin{center}
\leavevmode
\epsfbox[100 495 500 690]{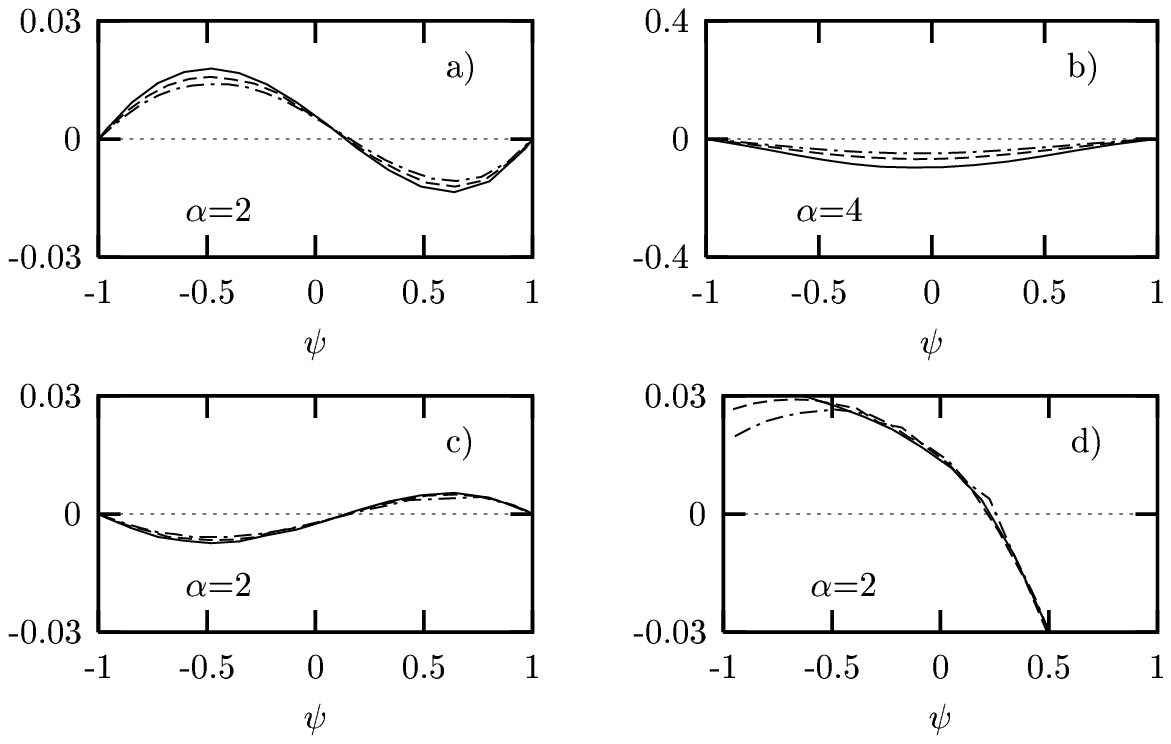}
\end{center}
\caption{
The ratio $R^T/k_F^\alpha$ of the seagull (a), pion-in-flight (b), 
vertex correlations (c) and 
self-energy (d) contributions to the transverse response and
$k_F^\alpha$ (with $\alpha$ as indicated in each panel) is plotted versus 
$\psi$ for $k_F$= 200 (solid), 250 (dashed) and
300 (dot-dashed) MeV/c. The momentum transfer is $q$=1 GeV/c.}
\end{figure}

\item In Fig.~6 we display the separate pion-in-flight and seagull
  contribution to $R^T$ for various values of $q$. It appears that the
  seagull term (dashed lines) is always larger than the 
  pion-in-flight term (solid lines), this dominance increasing with $q$ and
  again reflecting the spin nature of the photon-MEC interaction.
  Moreover whereas the pion-in-flight term is always negative, the
  seagull changes sign with $\omega$, inducing a (mild) softening of
  the response.

\item In Fig.~7a we study the evolution with $q$ of the MEC: it clearly
  appears that their relative contribution to $R^T$ decreases with $q$,
  but {\bf does not vanish for
  large values of q}.  This point is best illustrated by displaying
  the response versus the scaling variable~\cite{yscaling} 
  \be \psi \equiv
  \frac{\varepsilon_0-1}{\varepsilon_F-1} \ ,  \ee 
  using eq.~(\ref{neededforpsi}). Indeed the range
  $-1\leq\psi\leq 1$ will then be common to all responses, no matter
  what is the value of $q$.  In the figure we plot the ratio
  $\rho^T=R^T_{MEC}/R^T_{free}$ for $q$=0.5, 1, 1.5, 2, 2.5 and 3
  GeV/c: we see the relative MEC contribution decrease in going
  from 0.5 to 1 GeV/c, but then it rapidly saturates at or slightly above
  $q$=1 GeV/c, where its value stabilizes, typically around 10$\%$. Thus,
  one can see from these results that at momentum transfers above 1 GeV/c
  {\bf scaling of the first kind is satisfied} for the MEC contributions
  considered in this work.
  Note also that, for high $q$ the MEC almost vanish in the vicinity 
  of the QEP ($\psi$=0).

\item In Fig.~7b we investigate the $k_F$ dependence of the MEC
  contribution.  Importantly, the latter is seen to {\bf grow with}
  {\boldmath{$k_F$}}, in contrast with the free response which 
  decreases as $k_F^{-1}$.  This point is again best studied by
  displaying the response as a function of $\psi$, since the response
  region broadens with $k_F$.  In Fig.~7b the ratio $\rho^T$ is plotted
  versus $\psi$ at $q$=1 GeV/c for $k_F$ varying from 200 to 300
  MeV/c. It clearly appears that the relative MEC contribution grows
  with $k_F$ (attaining a value of about 20$\%$ at
  $k_F=$300 MeV/c).

  Indeed, our earlier studies of the current matrix elements
  \cite{Ama98} can be used in analyzing the functional dependence on $k_F$. 
  There a ``semi-relativistic'' approach was followed in which, 
  once the leading power dependence on the dimensionless variable $h/m$ was
  identified, where $h=|{\bf h}|$ is the three-momentum of any particle 
  below the Fermi surface, all higher-order terms in $h/m$ are neglected,
  since $h/m$ is bounded by $\eta_F=k_F/m$ and the latter is generally 
  quite small. For the matrix elements this was shown in \cite{Ama98} to 
  be an excellent approximation. Here we see a detailed verification of those 
  ideas. In fact, a full analysis at very small Fermi momentum
  shows that the seagull contribution has a leading power of 
  $k_F^2$, while the pion-in-flight contribution begins as $k_F^4$ --- 
  the former because it involves one- and two-body spin currents, 
  the latter because it involves convection currents and therefore 
  one higher power of momentum in both the one- and two-body contributions.
  Thus, if we divide the fully-relativistic results obtained in this work
  by these respective factors of $k_F^n$ with $n=2$, 4 for seagull and 
  pion-in-flight, respectively, we should see very little residual dependence
  on $k_F$, at least at very low Fermi momentum. 
  Indeed, even for values of the Fermi momentum that are typical of nuclei
  across the periodic table this is still roughly the case, as can be 
  seen by examining Fig.~8, panels $a$ and $b$. The relatively small 
  remaining spread with varying $k_F$ reflects the quality of the 
  semi-relativistic treatment in which only the leading $k_F$ dependence
  is taken into account. Of course, in the present work no approximation
  is made at all and it is not necessary to use the rough power dependence
  except as an orientation to assess the degree to which the two-body MEC 
  effects depend on the density in a manner that is different from the 
  behaviour seen in the one-body currents.

  It is also
  interesting to note at this juncture that, if we attempt to find the
  best choice for the power $n$ in the two cases, with $n$ free to be
  non-integer, specifically for nuclei in the range $200 < k_F < 260$
  (i.e., excluding only the lightest nuclei), then the respective powers
  turn out to be somewhat lower than 2 and 4, namely, about 1.5 and 2.5.
  
  Given that once the leading powers of $k_F$ are removed the 
  integrals remaining in the two cases are of roughly similar
  magnitude, one might expect the pion-in-flight contribution to be
  much smaller than the seagull contribution, at least at high $q$
  where such arguments can be made relatively easily (at low $q$ the
  pion propagators, etc. make the argument less secure). Indeed, having
  powers of $k_F^4$ and $k_F^2$, as discussed above, would imply that
  at high $q$ typically the seagull will win by roughly a factor of
  $(m/k_F)^2\cong 16$, and this is borne out in Fig.~6.

  Thus, overall the seagull contribution is dominant and so an overall 
  leading-$k_F$ dependence of about $k_F^2$ is appropriate. Since the RFG
  has a factor of $k_F^{-1}$ this means that the relative dependence --- 
  these two-body 1p-1h pionic contributions compared with the RFG --- goes 
  approximately as $k_F^3$. That is, when studying the scaling behavior of
  the second kind, these two-body MEC processes {\bf violate the 
  second-kind scaling by roughly three powers of} {\boldmath{$k_F$}}. 
  Clearly the effect is a rapid function of
  the Fermi momentum (or equivalently, of the density):
  for example, if one considers the cases $^2$H/$^4$He/heavy nuclei
  with Fermi momenta of approximately 55/200/260 MeV/c, respectively, 
  then the 1p-1h MEC contributions amount to 
  0.1/5/10\% of the total transverse response, respectively 
  (normalizing to 10\% for the heavy nucleus case --- see the first bullet).

\begin{figure}[hp]
\begin{center}
\leavevmode
\epsfbox[100 495 500 685]{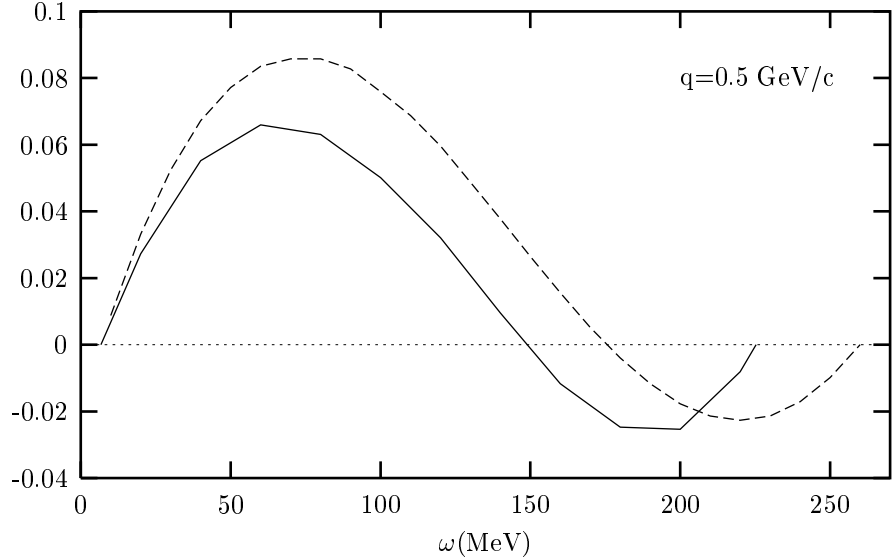}
\end{center}
\caption{
Transverse MEC contribution (solid) compared with the 
non-relativistic calculation of [15]
(dashed).
}
\end{figure}

\begin{figure}[hp]
\begin{center}
\leavevmode
\epsfbox[100 550 500 690]{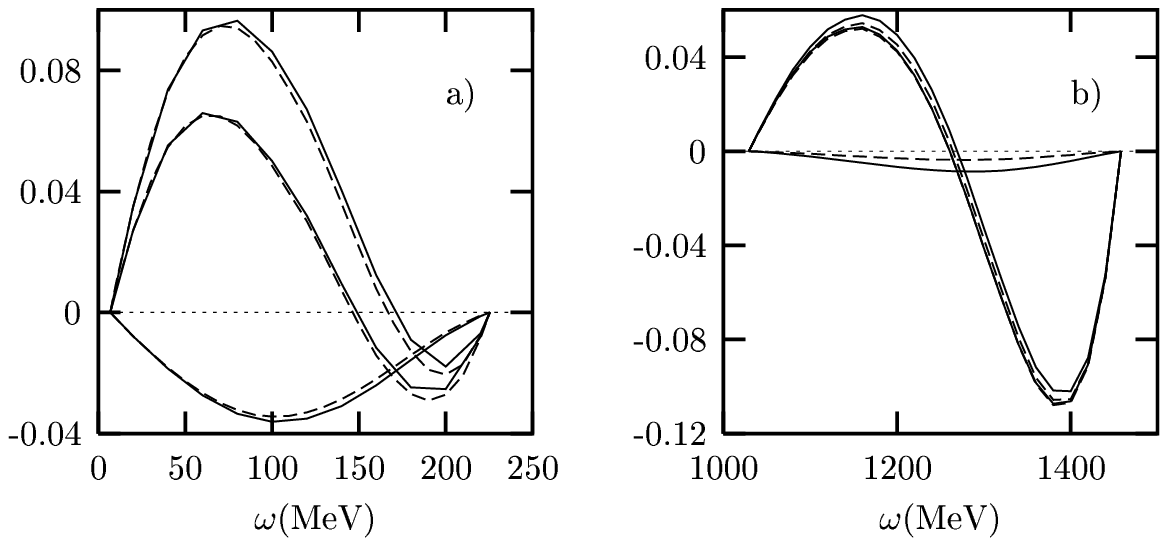}
\end{center}
\caption{
MEC contribution to $R^T$ versus $\omega$ with 
dynamic (solid curves) 
and static (dashed curves) pion propagator at $q$= 0.5 (a) and 2 (b) GeV/c. 
The separate pion-in-flight and seagull contributions 
are displayed.
}
\end{figure}

\item In Fig.~9 we compare the MEC contribution to the transverse
  response with the non-relativistic calculation of~\cite{Ama94b}
  for $q=500$ MeV/c. For this comparison we use $\Gamma_\pi=1$ and
  the static pion propagator in the relativistic calculation. The effect
  of static versus dynamic pion propagator will be discussed in the next
  item. From Fig.~9 we see that, apart from the difference stemming from the
  relativistic kinematics which shrinks the response domain, the
  relativistic responses are about $30\%$ smaller than 
  the non-relativistic ones, indicating that relativity
  plays an important role even for not so high $q$-values.

\item Finally the impact on the responses of the relativistic
  propagator $\Delta_\pi(K)=(K^2-m_\pi^2)^{-1}$ as compared to the
  static one $\Delta_\pi^{(n.r.)}({\bf k})=-({\bf k}^2+m_\pi^2)^{-1}$,
  which is commonly used in non-relativistic calculations, is
  explored.  In Fig.~10 the pion-in-flight, seagull and total MEC
  contributions to $R^T$ are evaluated for $q$=0.5 and 2 GeV/c using
  both propagators: it appears that the dynamical propagator affects
  the pion-in-flight more than the seagull term (it increases the 
  latter by more than a factor 2 at $q$=2 GeV/c); however, the two
  effects tend to cancel, so that their net effect is not very
  significant.

\end{itemize}

\begin{figure}[hp]
\begin{center}
\leavevmode
\epsfbox[100 495 500 690]{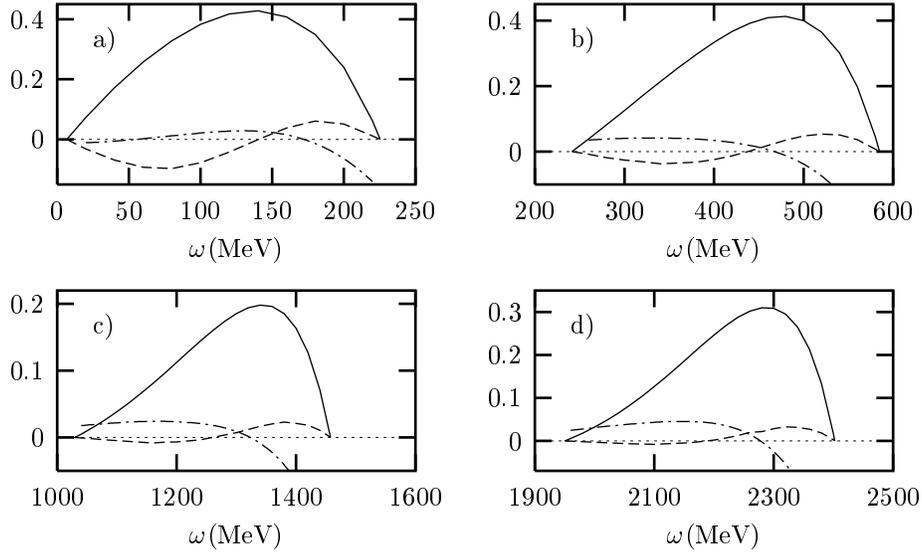}
\end{center}
\caption{
Correlation current contributions to $R^L$ versus $\omega$ at $q$=0.5 (a),
1 (b), 2 (c) and 3 (d) GeV/c. Solid: free; 
dot-dashed: self-energy; dashed: vertex correlations.
}
\end{figure}

\begin{figure}[hp]
\begin{center}
\leavevmode
\epsfbox[100 495 500 690]{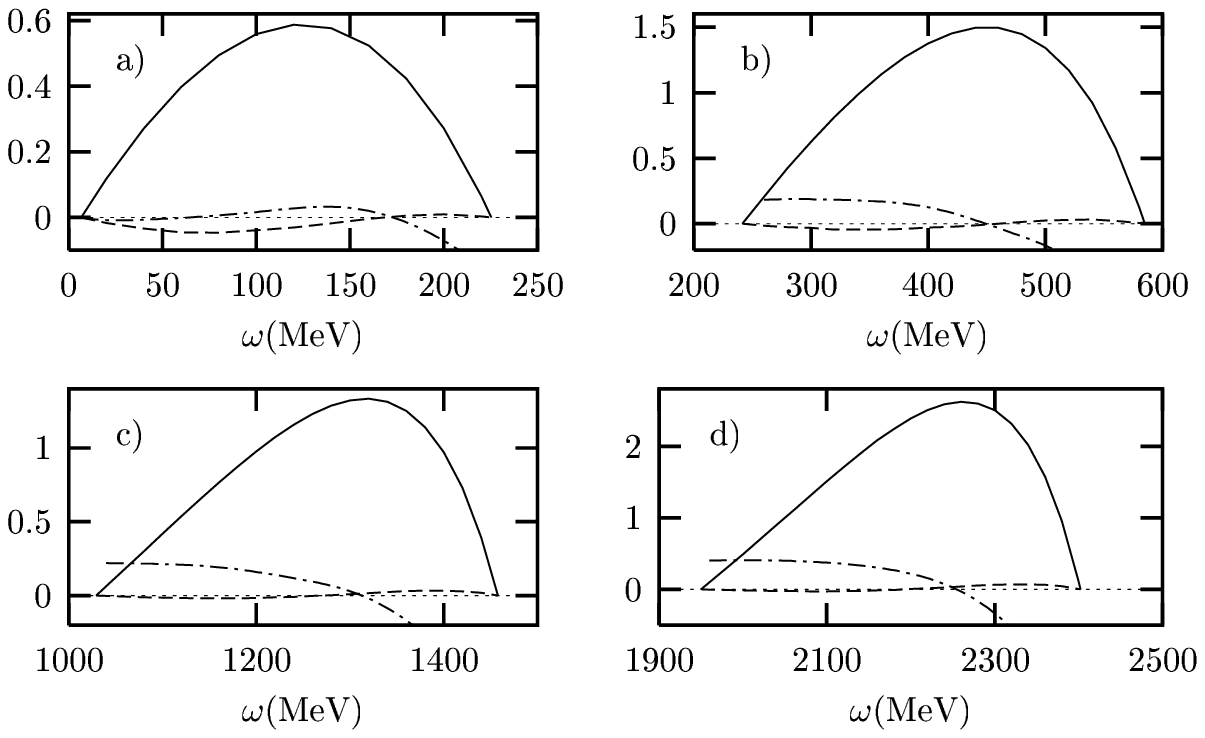}
\end{center}
\caption{
Correlation current contributions to $R^T$ versus $\omega$
at $q$=0.5 (a),
1 (b), 2 (c) and 3 (d) GeV/c.  Solid: free; 
dot-dashed: self-energy; dashed: vertex correlations.
}
\end{figure}

\subsection{Vertex correlations}
\label{sec:vc}

Where the v.c. are concerned we have found the following:

\begin{itemize}

\item
Their action, while substantial in both the longitudinal and 
transverse channel, is actually dominant in the former by roughly a
factor of 3:1,
as is evident from Figs.~11 and 12, where the v.c. contribution 
(dashed) displayed for
several values of $q$ together with the free responses (solid).
This outcome relates to the minor role played by the isoscalar contribution
in the transverse response, in turn due to the smallness of the isoscalar
magnetic moment.
Worth pointing out is the oscillatory behavior versus $\omega$ of the vertex
correlations, which induces a hardening of the responses.

In addition the seagull and vertex correlations tend to cancel 
in the transverse channel, especially for low values of
$q$, whereas for higher $q$ the MEC dominate (see Fig.~13).
Note that both the seagull and v.c. exactly vanish at the same value of 
$\omega$, the latter coinciding with the QEP for high momentum
transfers.

\begin{figure}[hp]
\begin{center}
\leavevmode
\epsfbox[100 495 500 690]{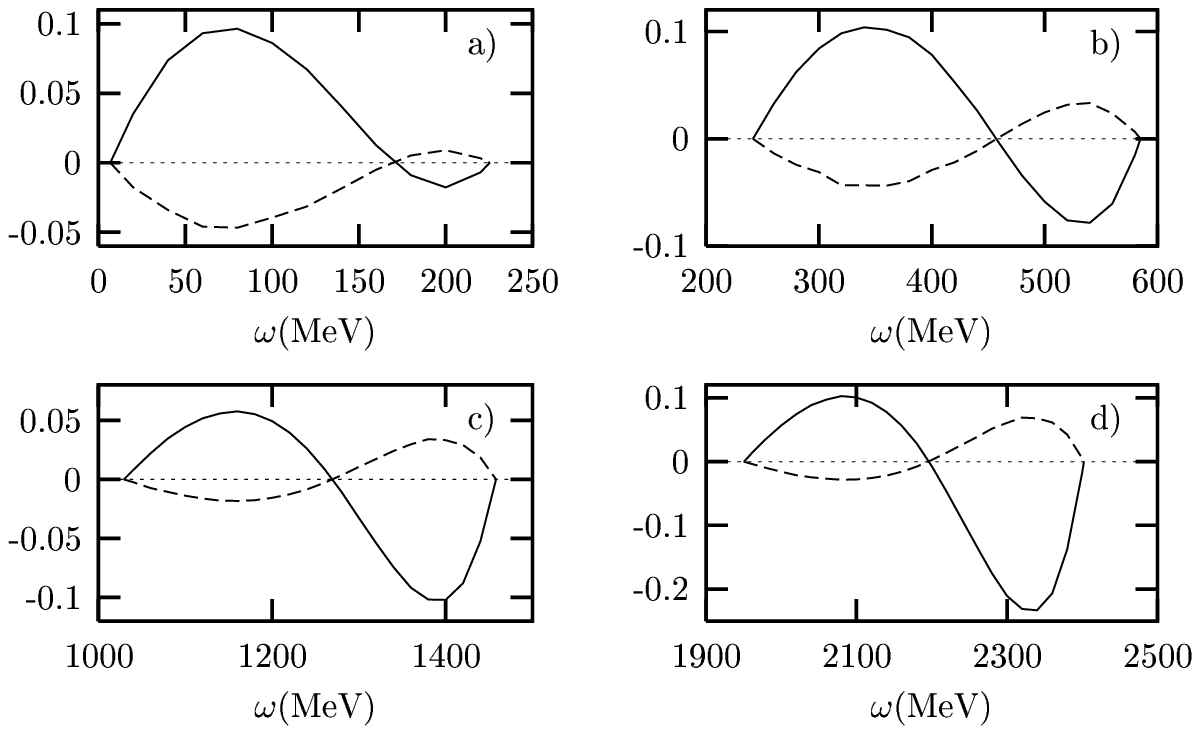}
\end{center}
\caption{
Seagull (solid) and v.c. (dashed) contributions to $R^T$ plotted 
versus $\omega$ at $q$=0.5 (a),
1 (b), 2 (c) and 3 (d) GeV/c.
}
\end{figure}

\begin{figure}[hp]
\begin{center}
\leavevmode
\epsfbox[100 555 500 690]{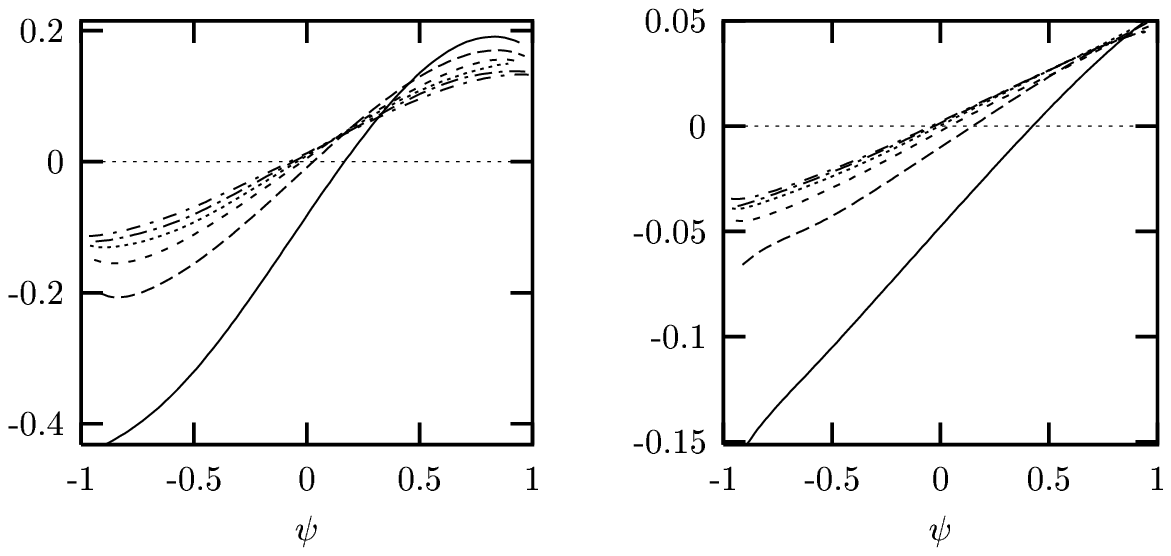}
\end{center}
\caption{
The ratio (v.c.)/free in the longitudinal (left panel)
and transverse (right panel) channel plotted versus $\psi$ at $k_F$=237 MeV/c
for various $q$. The curves are labelled as in Fig.~7a.
}
\end{figure}

\item
In Fig.~14 we study the evolution with $q$ of the v.c. in the
longitudinal and transverse channel respectively plotting the ratios
$\rho^{L(T)}=R^{L(T)}_{v.c.}/R^{L(T)}_{free}$ 
at $q$=0.5, 1, 1.5, 2, 2.5 and 3 GeV/c.
Clearly the v.c. do not saturate quite as rapidly as the MEC, although
their behaviour is rather similar and saturation again occurs somewhere
above $q=1$--1.5 GeV/c: thus again {\bf scaling of the first kind is
achieved} at high momentum transfers for these contributions.
Moreover, similarly to the MEC case, for high $q$ the v.c. almost vanish 
around the QEP ($\psi$=0). In fact the v.c. contributions and the seagull MEC 
contributions are rather similar in shape, but opposite in sign (cf.
Figs.~7a and 14 (right panel)). Thus, they tend to cancel and, since the
relative magnitude of the seagull to the v.c. is roughly 2:1, the net effect
of the two together is to behave the way the former does but with roughly
1/2 its strength.

\begin{figure}[hp]
\begin{center}
\leavevmode
\epsfbox[100 550 500 690]{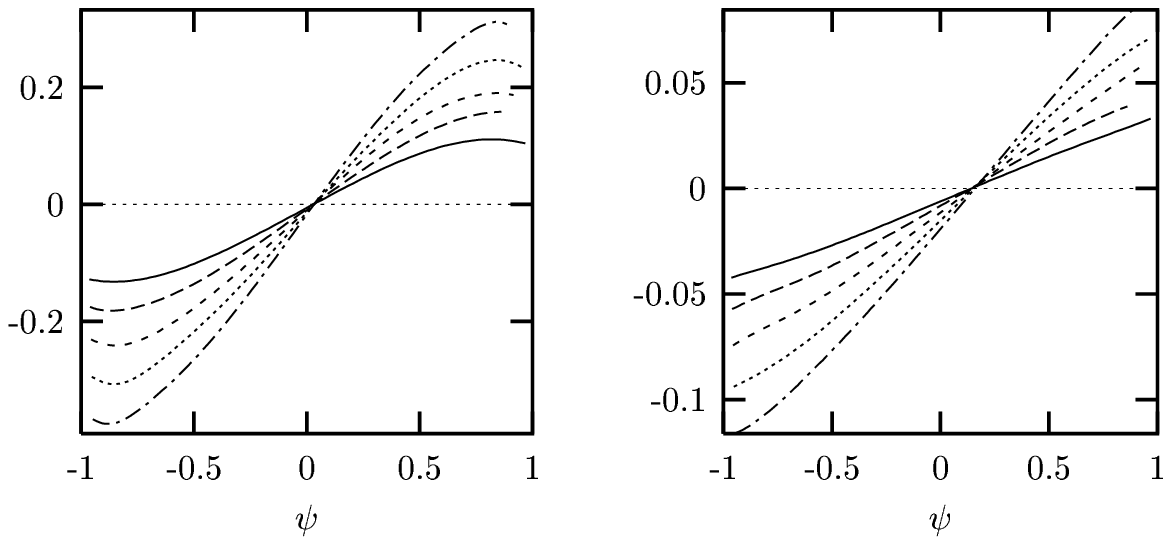}
\end{center}
\caption{
The ratio (v.c.)/free in the longitudinal (left panel)
and transverse (right panel) channel plotted versus $\psi$ at $q$=1 GeV/c
for various $k_F$. The curves are labelled as in Fig.~7b.
}
\end{figure}

\begin{figure}[hp]
\begin{center}
\leavevmode
\epsfbox[100 495 500 685]{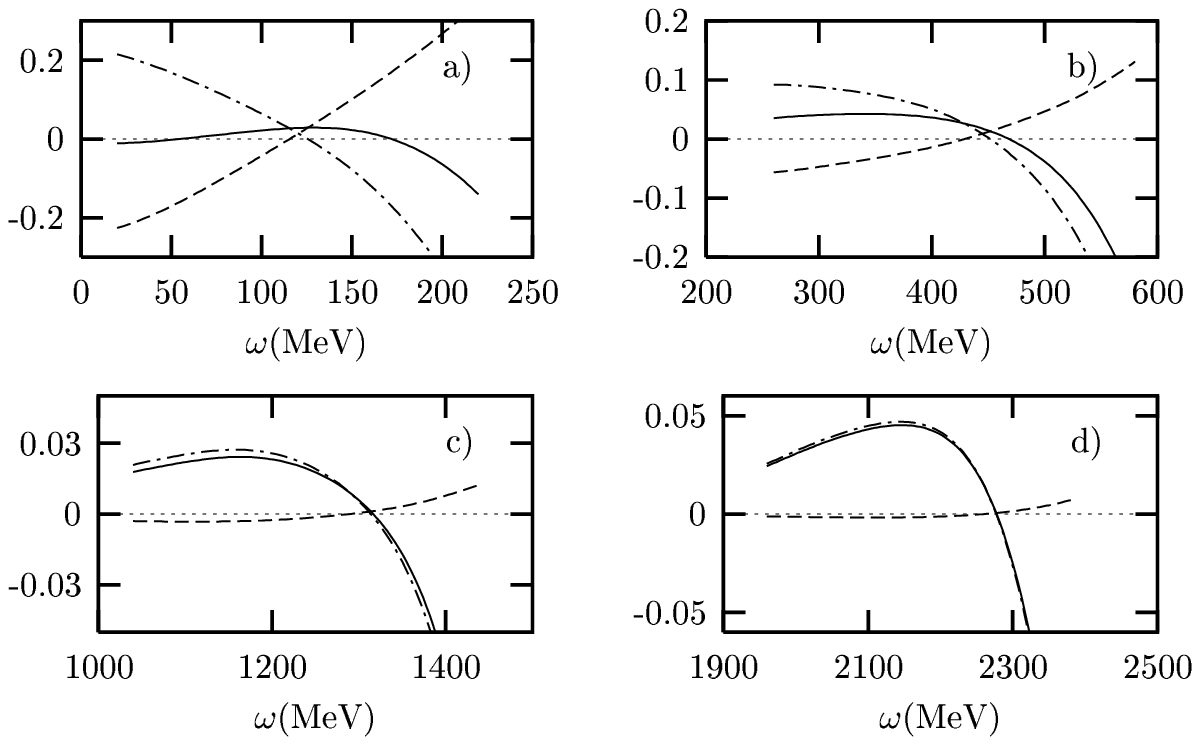}
\end{center}
\caption{
Particle (dashed) and hole (dot-dashed) contributions to the longitudinal
self-energy (solid) versus $\omega$ for $q=0.5$ (a), 1 (b), 1.5 (c)
and 2 (d) GeV/c.
}
\end{figure}

\item
In Fig.~15 we explore the $k_F$ dependence of the vertex correlations.
They are found to grow with $k_F$, much as the MEC do. Again from
a semi-relativistic point of view, expanding only in powers of hole
three-momenta over nucleon mass, we find a leading behaviour for $k_F$
small that goes as $k_F^2$. 
In Fig.~15 the ratios $\rho^L$ and $\rho^T$ are plotted versus 
$\psi$ at $q$=1 GeV/c for $k_F$ varying from 200 to 300 MeV/c and we clearly
see that the v.c. contribution grows with $k_F$. As in the MEC seagull
case, for Fermi momenta in the typical range (200--250 MeV/c) a
slightly lower power of $k_F^{3/2}$ actually provides the best fit
to the full results. The range seen in Fig.~8c reflects the (weak)
dependence on $k_F$ of terms ignored in the semi-relativistic
approximation. The basic conclusion is similar to that made above
for the seagull contribution and hence the total MEC at high $q$, namely,
{\bf scaling of the second kind is badly broken} by effects that go 
roughly as $k_F^3$.

\end{itemize}

\subsection{Self energy}
\label{sec:se}

We have found that the self-energy contribution
results from a quite delicate cancellation
between the responses having only the particle or only the hole 
dressed, as shown in Fig.~16 in the 
longitudinal channel for $q$=0.5, 1, 2 and 3 GeV/c
(similar results hold in the transverse case). This was
already pointed out in~\cite{Barb93} within the framework of a 
treatment in which relativistic effects were partially incorporated,
and it is now confirmed here in a fully relativistic context.

Whereas this cancellation is very substantial at $q$=0.5 GeV/c, 
as the momentum
transfer increases the imbalance between the two contributions grows.
Indeed the response associated with the particle self-energy is suppressed by
the form factors
and by the pion propagator, but that coming from the hole self-energy is
not. As a result, for $q\geq$ 2 GeV/c the total self-energy response
almost coincides with the hole result alone and induces a moderate softening
to the free response. 
Note that the s.e. contribution does not vanish on the borders of the 
response region. Moreover
for high values of $\omega$ (close to the upper border) it becomes 
very large (see Figs.~11 and 12) and yields a vanishing 
(or even negative) response, clearly pointing to the 
insufficiency of a first-order perturbative treatment in this 
kinematical region. This effect
was already present in the partially relativized analysis of~\cite{Barb93} 
and, significantly, it is emphasized by our fully relativistic calculation.
Therefore the summation of the full Fock series becomes 
necessary in this case.

\begin{figure}[hp]
\begin{center}
\leavevmode
\epsfbox[100 495 500 685]{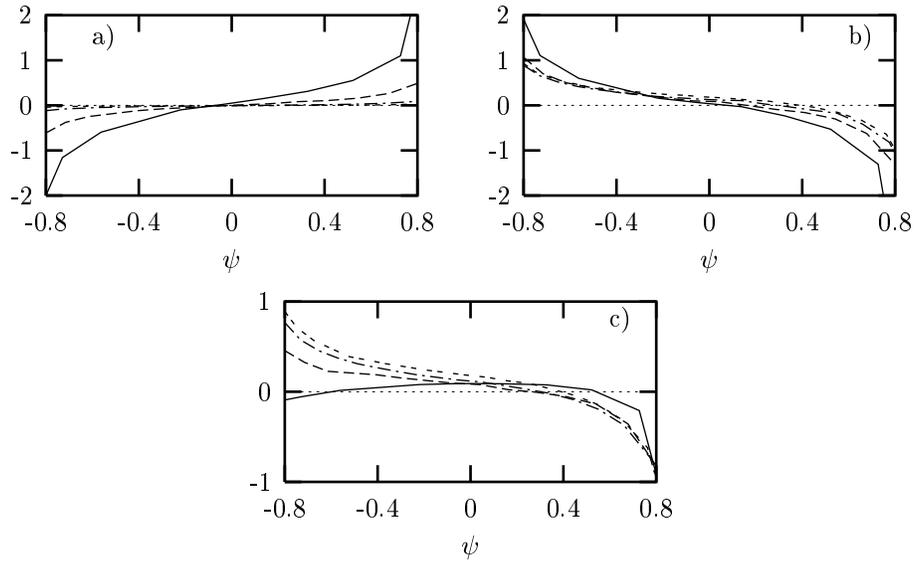}
\end{center}
\caption{
The ratio of the particle (a), hole (b) and total (c) s.e. contribution 
to the free longitudinal response is plotted versus $\psi$ for 
$q$=0.5 (solid), 1 (dashed), 2 (dot-dashed) and 3 (short-dashed) GeV/c.
}
\end{figure}

\begin{figure}[hp]
\begin{center}
\leavevmode
\epsfbox[100 495 500 685]{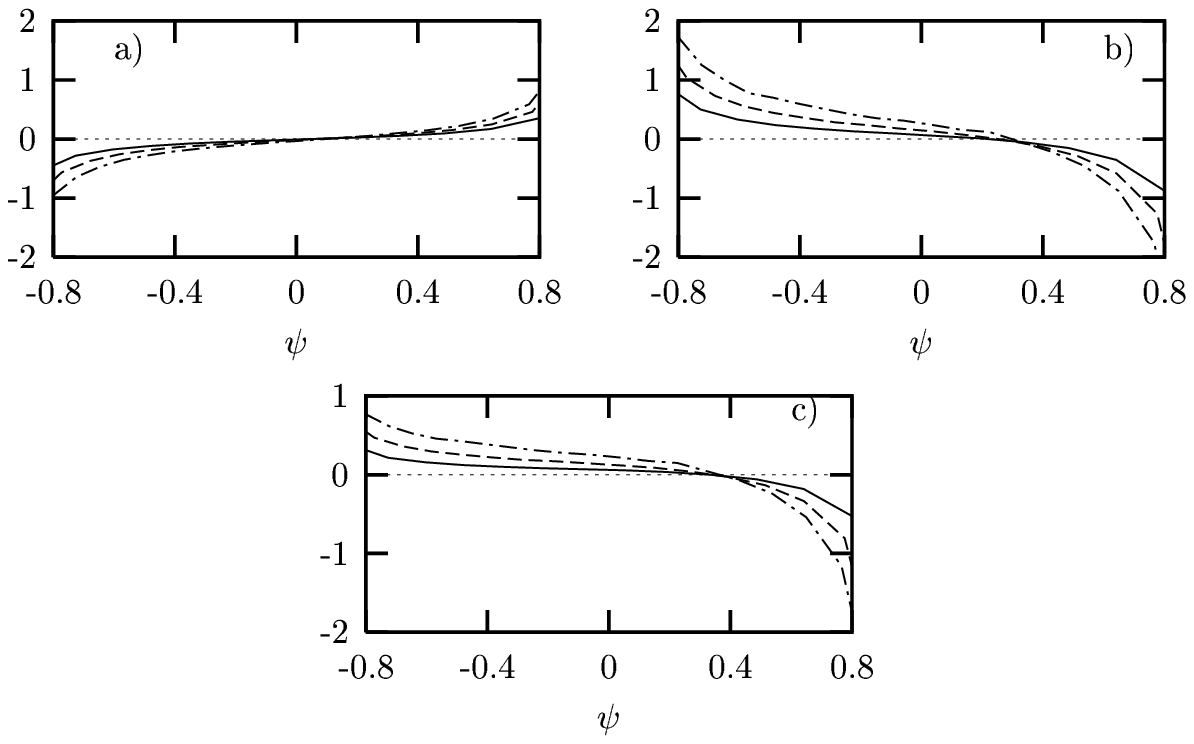}
\end{center}
\caption{
The ratio of the particle (a), hole (b) and total (c) s.e. contribution 
to the free longitudinal response is plotted versus $\psi$ for 
$k_F$=200 (solid), 250 (dashed) and 300 (dot-dashed) MeV/c and $q$=1 GeV/c.
}
\end{figure}

Finally we explore the scaling and superscaling properties of the self-energy
correlations by studying their momentum transfer- and density-dependence.
In Fig.~17 we plot the ratio $\rho^L=R^L_{s.e.}/R^L_{free}$ versus $\psi$ for
the particle (panel a), hole (panel b) and total (panel c) self-energies at 
$q$=0.5, 1, 2 and 3 GeV/c.
As expected, the particle contribution decreases with $q$, going to zero
at $q\simeq 2$ GeV/c, whereas the hole contribution, although 
also decreasing with $q$ when not too high, saturates
for $q\geq$1 GeV/c. As a result the total self-energy, displayed in
panel (c), grows with $q$ (in contrast to all other cases considered above)
in the range $q$=0.5-2 GeV/c,
then stabilizes typically at about 30-40\% of the free response 
to the left of the QEP, thus inducing the 
above-mentioned softening of the longitudinal response.
Similar results are found in the transverse channel. In summary, again
{\bf scaling of the first kind is achieved} at momentum transfers somewhat
below 2 GeV/c.

In Fig.~18 we display the same ratio for three different
values of the Fermi momentum at $q$=1 GeV/c. 
It appears that, as for the other two-body correlations, the 
self-energy relative contribution grows with $k_F$, specifically
$R^L_{s.e.} \propto k_F^2$, although not uniformly in $\psi$
(see Fig.~8, panel $d$) --- recall that in the first-order analysis presented
in this paper the edges of the response region are not treated 
adequately for the self-energy contribution and thus should not be
taken too seriously. Where the self-energy contribution is correctly
modeled (away from the edges) we again see {\bf breaking of second-kind
scaling} by roughly $k_F^3$.

\vspace{1cm}

\begin{figure}[hp]
\begin{center}
\leavevmode
\epsfbox[100 495 500 685]{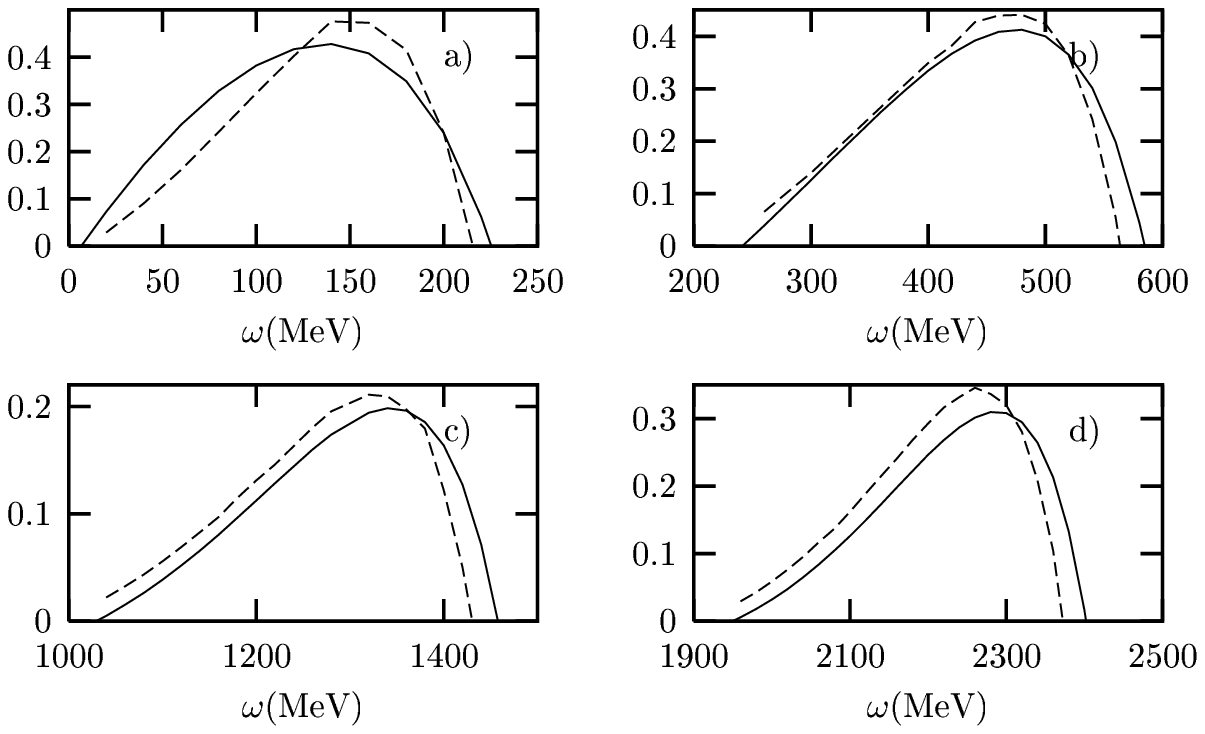}
\end{center}
\caption{
Longitudinal response versus $\omega$ including all first-order contributions
(dashed) compared with the free result (solid) at $q$=0.5 (a), 1 (b),
2 (c) and 3 (d) GeV/c.
}
\end{figure}

\begin{figure}[ht]
\begin{center}
\leavevmode
\epsfbox[100 495 500 685]{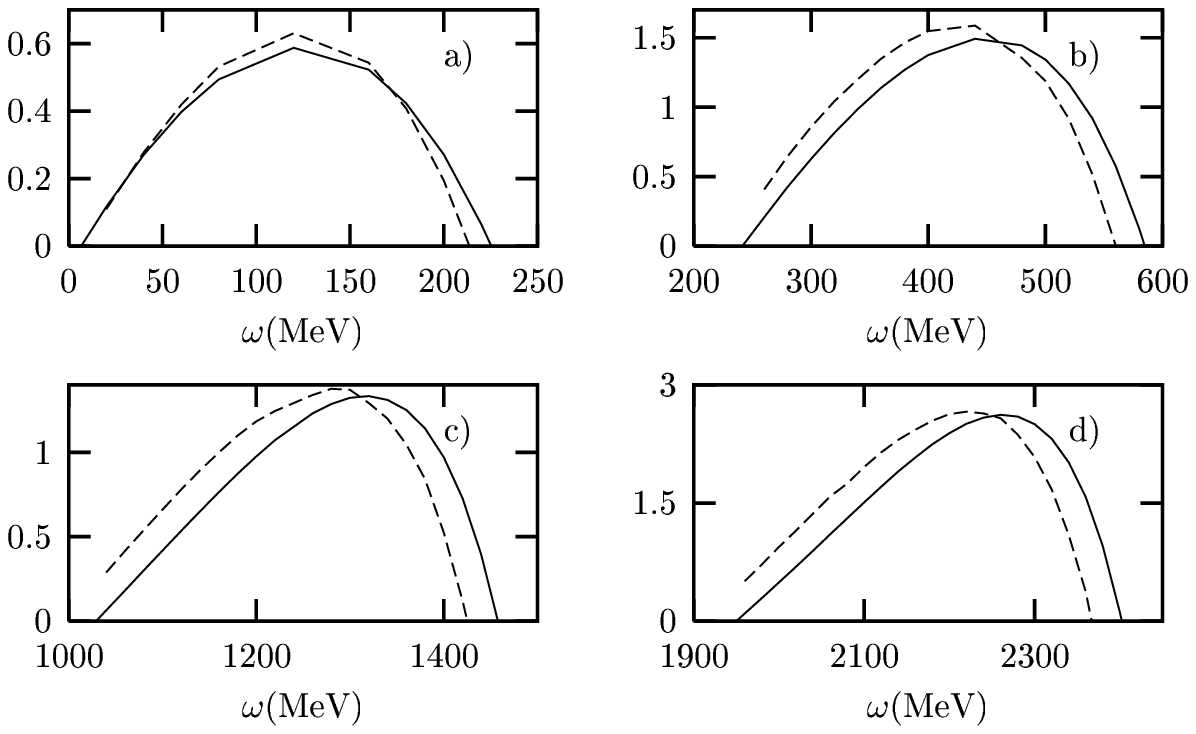}
\end{center}
\caption{
Transverse response versus $\omega$ including all first-order contributions
(dashed) compared with the free result (solid) at $q$=0.5 (a), 1 (b),
2 (c) and 3 (d) GeV/c.
}
\end{figure}

To complete the presentation of our results we display in Figs.~19 and 20
the global responses in first order of perturbation theory and compare them
with the zeroth order ones (free responses) for several momentum transfers.
Here one assesses the impact of the global two-body 
current contribution to the responses.
First the overall effect of the two-body currents appears 
sufficiently modest to justify our first-order treatment. 
Next the softening at large $q$ appears to be common to both 
L and T channels,
whereas at low $q$ the longitudinal response displays a hardening 
that is absent in the transverse one.
Also evident is the already-noted almost vanishing of 
the two-body correlation contribution at the peak of the free responses.
Finally the unrealistic dominance of the self-energy 
contribution on the upper
border is apparent.

\section{Conclusions}
\label{sec:concl}

In this paper we have addressed some of the ingredients that enter when
attempting to model the longitudinal and transverse nuclear response 
functions for inclusive quasielastic electron scattering.
Because of its complexity this problem cannot presently be viewed as solved:
indeed, although many papers claim success in accounting for the data, a
closer scrutiny reveals a different situation. It is 
not only that contributions left out in various
analyses are far from being small, but, even more serious, fundamental
physics principles (Lorentz covariance, gauge invariance and unitarity)
turn out patently to be violated.
Thus the successes in reproducing the experiments, regrettably few both in
connection with the L/T separation and with the set of nuclei or range
of $q$ explored, often reflect more an adjusting of parameters than a
real understanding of the physics involved in the QEP.

Searching for an improvement of this situation we have engaged in a more
systematic approach to the problem, assuming as a zeroth-order approximation
the RFG model. This appears justified on two counts:
i) because RFG is Lorentz covariant,
ii) because the surface of the nucleus is largely irrelevant in the QEP.

One of our basic themes has been the implementation of Lorentz covariance.
In past work~\cite{Ama98} we attempted to approximate the full theory
by identifying a dimensionless variable that is 
small enough to be suitable to use in setting up a non-relativistic 
expansion of the responses (namely the momentum of a nucleon lying below
the Fermi surface compared with its mass). In contrast, in the present work 
no expansion whatsoever is involved and our calculation fully respects 
relativity. The motivation for this should be clear: 
the theoretical analyses of
the present and future experiments at TJNAF demand as full an implementation
of relativity as is feasible.

Since the physics of the QEP is remote not only 
  from the physical surface of the nucleus, but from the Fermi surface as well,
we consider a treatment in terms of nucleonic and mesonic degrees of freedom
(the latter viewed both as force and current carriers) to be appropriate.
Hence, as a first step, we focus on pions as they can be expected under
these conditions to be an important, if
not the major, carrier of the currents that respond to an external
electromagnetic field impinging on the nucleus.
The pions, in our framework, are dealt with in first order of perturbation
theory, since their effects on the free responses 
of the RFG are not expected to be too disruptive.

A point worth making in connection with all of the two-body currents
considered in the present work is the following: 
their contributions appear to be quite small
(if not vanishing) at the peak of the quasielastic responses. This 
corresponds to viewing the Fermi sphere to be 
split into two domains, an inner one
associated with a sphere with a radius smaller than $k_F$, and the
other corresponding to the spherical shell representing 
the difference between
the two concentric spheres. The contributions tend to cancel one another
and at the peak, where the whole Fermi sphere is involved in the response, 
they nearly do so completely. Of course the radius of the inner sphere 
and the sign of the contributions are current dependent.


Gauge invariance is a fundamental property we have also addressed in this
work. In particular we have verified whether or not the continuity equation
it implies is valid order by order in perturbation theory.
Notably we have succeeded in showing that the continuity equation for the
one-body (single-nucleon) and the two-body (MEC and correlations) currents
{\em is} indeed fulfilled; hence our approach deals consistently with forces
and currents. At the same time we have also shown how crucial relativity
is in achieving this and how much care is needed in 
dealing with the ph
matrix element of the self-energy current.
Actually the present approach points to the necessity of renormalizing both
the energies and the spinors in the self-energy ph matrix elements.
Work under these lines is under progress \cite{Ama01}.

As far as the self-energy contribution is concerned, the present 
results show that for high momentum transfers a first-order
perturbative treatment seems to be insufficient to deal with the
kinematical region corresponding to high values of $\omega$ (close to
the upper border of the response region).  This shortcoming demands
the inclusion of higher-order perturbative terms, namely the summation
of the whole Fock series, a task performed in a partially relativized
framework~\cite{Barb93} but not yet achieved in a fully relativistic
one.  It might be worth pointing out that the above shortcoming relates to the
``wrong'' analytic properties of our first-order
polarization propagator: it should be
a meromorphic function of the energy, yet it displays a double pole.

With regard to the other contributions of the two-body currents to $R^L$ and
$R^T$, our results are highly satisfactory. Indeed for the MEC we have found
that their contribution is small enough to be well handled in first order.
In particular both the pion-in-flight and seagull contributions are very
small in the L channel where the virtual photon exchanged 
between the electron
and the Fermi gas couples to the charge of the pion. We thus see that the MEC
affect the Coulomb sum rule very marginally.
In the T channel instead the MEC are more 
significant and the following features
emerge (see also Section~\ref{sec:res} for itemized observations):
\begin{itemize}
\item  the seagull contribution dominates, in accord with the spin nature
of the MEC physics, and displays an oscillating behaviour versus $\omega$
yielding a (modest) softening of the response;
\item  the MEC contribution does not vanish when $q$ increases, 
but ultimately at very high $q$ it displays scaling of the first kind;
\item the MEC contributions break the scaling of the second kind; i.e.,
they go as higher powers of $k_F$ than does the RFG. In fact, the total
MEC contribution goes roughly as $k_F^3$ with respect to the RFG and thus
when $k_F$ becomes small the MEC contribution tends to vanish.
Hence, in going from the lightest nuclei where $k_F$ is as small as
about 55 MeV/c (viz., for the deuteron) to heavy nuclei where it may be 
as large as 260 MeV/c, one should see the 1p-1h quasielastic MEC 
contribution go from being essentially negligible to becoming typically 
5--10\% of the total transverse response. 

\end{itemize}

With regard to the correlation contribution arising 
from the vertex corrections
(v.c.), we have found that the L channel dominates over the T channel
by a margin of roughly 3:1. The longitudinal response effectively picks up
only these correlation contributions, since the MEC effects are so
small there, and the former contribute to the total at roughly the
10--15\% level. Indeed, were these to be the only contributions needed
in addition to the RFG response itself, then we would expect the total
to shift in $\omega$ (or in $\psi$). Since the correlation 
contributions are roughly symmetrical about the QEP,
it is important to note that 
their impact on the Coulomb sum rule should be very small, perhaps only at the
few percent level.

The correlation contribution
to $R^T$ is similar to the MEC contribution, but is smaller, 
roughly 1/2 the size of the latter. Since the two are of opposite sign
they tend to cancel and thus the total is similar to the MEC contribution
but is cut down by a factor of two. This statement is of course
only approximate, since the correlation contribution
attains its asymptotic value as a function of $q$ 
somewhat less rapidly than does the MEC contribution. 
Interestingly, both the seagull and v.c.
contributions appear to vanish at exactly the same value of $\omega$.
Overall the total (the sum of 1p-1h MEC + 1p-1h correlations)
yields contributions to be added to the RFG that (1) do not go away
as $q$ becomes very large, (2) attain scaling of the first kind at 
momentum transfers somewhat above 1 GeV/c, while (3) they clearly violate
scaling of the second kind, going approximately as $k_F^3$ relative
to the RFG result (which of course scales).

In conclusion, in this paper we have pursued the program of a systematic 
analysis of the inclusive nuclear responses in the domain of the QEP and
beyond. The motivation for this effort lies in the recognition that in
this energy domain, unlike for the ground and low-lying excited states of
nuclei, the interaction among the constituents of the system cannot be 
subsumed into a potential. Currents are important; and moreover
the consistent treatment of currents and forces is important. For
the first time one is able to investigate the scaling behavior of 
at least some of the contributions to the inclusive cross section 
under high-energy conditions using a non-trivial
(interacting) model in which issues relating to covariance and 
gauge invariance are addressed consistently.

\subsection*{Acknowledgements}
This work was partially supported by funds provided by DGICYT (Spain) 
under Contracts Nos. PB/98-1111, PB/98-0676 and PB/98-1367 and the Junta
de Andaluc\'{\i}a (Spain), by the Spanish-Italian Research Agreement
HI1998-0241, by the ``Bruno Rossi'' INFN-CTP Agreement, by the 
INFN-CICYT exchange and in part by the U.S. Department of Energy under
Cooperative Research Agreement No. DE-FC02-94ER40818.

\subsection*{Appendix A. The gauge invariance of the two-body
                         currents ph matrix elements}

Following the study presented in Section~\ref{sec:Form} 
of gauge invariance at the
level of the free-space particle-particle matrix elements, here
we extend the analysis to the particle-hole channel, deriving the
contribution to the continuity equation of the isoscalar and isovector 
s.e., v.c. and MEC particle-hole matrix elements.
We start by evaluating the divergence of the
correlation particle-hole matrix element 
$\langle ph^{-1}|{\hat j}_\mu ^C|F\rangle$ 
for the s.e. and v.c. contributions;
next we address the MEC ph matrix elements.

\begin{itemize}
\item
\underline{Self energy}
\end{itemize}
\noindent
   From eqs.~(\ref{Sp},\ref{Sh}) we get 
\ba
Q\cdot {\cal H}_p =
-\frac{3f^2}{2m V^2m_\pi^2} \frac{m}{\sqrt{E_\np E_\nh}} 
\sum_{\nk}\frac{m}{E_\nk}
\overline{u}(\np) 
 \frac{(\Pbar-\Kbar) (\Kbar - m) (\Pbar-\Kbar)}{(P-K)^2-m_\pi^2} 
              S_F(P) F_1 \Pbar u(\nh) \nonumber\\
Q\cdot{\cal H}_h=
-\frac{3f^2}{2m V^2m_\pi^2} \frac{m}{\sqrt{E_\np E_\nh}} 
\sum_{\nk}\frac{m}{E_\nk}
\overline{u}(\np) 
             F_1 \Qbar S_F(H)
              \frac{(\Kbar-\Hslash)(\Kbar - m)(\Kbar-\Hslash)}
{(K-H)^2-m_\pi^2} u(\nh) \ .\nonumber
\ea
Note that $F_1$ cannot be taken out of the matrix element since it acts 
on the isospinors. Now from the relations
\ba
S_F(P) \Qbar u(\nh) &=& u(\nh)
\\
\overline{u}(\np) \Qbar S_F(H) &=& - \overline{u}(\np) 
\\
\overline{u}(\np) (\Pbar-\Kbar)(\Kbar - m) &=& 2m \,\overline{u}(\np) 
(\Pbar-\Kbar) 
\\
(\Kbar - m) (\Kbar-\Hslash) u(\nh) &=& -2m (\Kbar - m)u(\nh) 
\ea
the following expressions are derived:
\ba
Q\cdot{\cal H}_p=
-\frac{3f^2}{V^2m_\pi^2} \frac{m}{\sqrt{E_\np E_\nh}} 
\sum_{\nk}\frac{m}{E_\nk}
\overline{u}(\np) 
 \frac{(\Kbar - m) (\Pbar-\Kbar)}{(P-K)^2-m_\pi^2}  F_1 u(\nh)
\label{Spslash} \\
Q\cdot{\cal H}_h=
-\frac{3f^2}{V^2m_\pi^2} \frac{m}{\sqrt{E_\np E_\nh}} 
\sum_{\nk}\frac{m}{E_\nk}
\overline{u}(\np) F_1 
              \frac{(\Kbar-\Hslash) (\Kbar - m)}{(K-H)^2-m_\pi^2} u(\nh) \ .
\label{SHslash}
\ea

\begin{itemize}
\item
\underline{Vertex correlations}
\end{itemize}
\noindent
  From eqs.~(\ref{F},\ref{B}) the four-divergence of the v.c. matrix element 
is found to be

\ba
Q\cdot{\cal F} 
=  -\frac{f^2}{V^2m_\pi^2} \frac{m}{\sqrt{E_\np E_\nh}}
\sum_{\nk}\frac{m}{E_\nk}
\overline{u}(\np) \gamma_5(\Kbar-\Hslash)
              S_F(K+Q)\tau_a F_1\tau_a \Qbar \gamma_5
              \frac{\Kbar-m}{(K-H)^2-m_\pi^2} u(\nh)
\nonumber\\
Q\cdot{\cal B}
= -\frac{f^2}{V^2m_\pi^2} \frac{m}{\sqrt{E_{\np} E_\nh}} 
\sum_{\nk}\frac{m}{E_\nk}
\overline{u}(\np) 
              \frac{\Kbar-m}{(P-K)^2-m_\pi^2}  
            \tau_a F_1 \tau_a \gamma_5 \Qbar S_F(K-Q) \gamma_5
              (\Pbar-\Kbar)u(\nh) .
\nonumber\ea
We now exploit the identities
\ba
S_F(K+Q) \Qbar (\Kbar + m) &=& +(\Kbar + m)
\\
(\Kbar + m) \Qbar S_F(K-Q) &=& -(\Kbar + m)
\label{ident}
\ea
to get finally 
\ba
Q\cdot{\cal F}
=            \frac{f^2}{V^2m_\pi^2} \frac{m}{\sqrt{E_\np E_\nh}}
 \sum_{\nk}\frac{m}{E_\nk}
\overline{u}(\np)  \tau_a F_1\tau_a 
 \frac{(\Kbar-\Hslash)(\Kbar-m)}{(K-H)^2-m_\pi^2} u(\nh)
\label{Fslash}
\\
Q\cdot{\cal B}
=            \frac{f^2}{V^2m_\pi^2}  \frac{m}{\sqrt{E_\np E_\nh}}
\sum_{\nk}\frac{m}{E_\nk}
\overline{u}(\np) \tau_a F_1 \tau_a
              \frac{(\Kbar - m)(\Pbar-\Kbar)}{(P-K)^2-m_\pi^2} 
              u(\nh) .
\label{Bslash}
\ea

If the expressions 
(\ref{Spslash},\ref{SHslash},\ref{Fslash},\ref{Bslash})
are split into their isoscalar and isovector parts, as illustrated in 
Section~\ref{sec:Pion}, we get
\ba
Q\cdot{\cal H}^{(S)}_p &=&
-\frac{3f^2}{V^2m_\pi^2} F_1^{(S)} \frac{m}{\sqrt{E_\np E_\nh}}
\sum_{\nk}\frac{m}{E_\nk}
\overline{u}(\np) 
 \frac{(\Kbar - m)(\Pbar-\Kbar)}{(P-K)^2-m_\pi^2} u(\nh)
\nonumber\\
Q\cdot{\cal H}^{(V)}_p &=&
-\frac{3f^2}{V^2m_\pi^2} F_1^{(V)} \frac{m}{\sqrt{E_\np E_\nh}}
\sum_{\nk}\frac{m}{E_\nk}
\overline{u}(\np) 
 \frac{(\Kbar - m) (\Pbar-\Kbar)\tau_3}{(P-K)^2-m_\pi^2} u(\nh)
\nonumber\\
Q\cdot{\cal H}^{(S)}_h &=&
-\frac{3f^2}{V^2m_\pi^2}  F_1^{(S)} \frac{m}{\sqrt{E_\np E_\nh}}
\sum_{\nk}\frac{m}{E_\nk}
\overline{u}(\np) 
\frac{(\Kbar-\Hslash)(\Kbar - m)}{(K-H)^2-m_\pi^2} u(\nh) 
\nonumber\\
Q\cdot{\cal H}^{(V)}_h &=&
-\frac{3f^2}{V^2m_\pi^2}  F_1^{(V)} \frac{m}{\sqrt{E_\np E_\nh}}
\sum_{\nk}\frac{m}{E_\nk}
\overline{u}(\np) 
\frac{(\Kbar-\Hslash)(\Kbar - m)\tau_3}{(K-H)^2-m_\pi^2} u(\nh) 
\nonumber\\
Q\cdot{\cal F}^{(S)}
&=&            +\frac{3 f^2}{V^2m_\pi^2} F_1^{(S)}\frac{m}
{\sqrt{E_\np E_\nh}}
 \sum_{\nk}\frac{m}{E_\nk}
\overline{u}(\np)
\frac{(\Kbar-\Hslash)(\Kbar - m)}{(K-H)^2-m_\pi^2} u(\nh)
\nonumber\\
Q\cdot{\cal F}^{(V)}
&=&            +\frac{ f^2}{V^2m_\pi^2} F_1^{(V)}\frac{m}{\sqrt{E_\np E_\nh}}
\sum_{\nk}\frac{m}{E_\nk}
\overline{u}(\np) 
\frac{(\Kbar-\Hslash)(\Kbar - m)}{(K-H)^2-m_\pi^2}
(\tau_3+i\varepsilon_{3ab}\tau_a\tau_b) u(\nh)
\nonumber\\
Q\cdot{\cal B}^{(S)}
&=&            +\frac{3f^2}{V^2m_\pi^2} F_1^{(S)} 
\frac{m}{\sqrt{E_\np E_\nh}}
 \sum_{\nk}\frac{m}{E_\nk}
\overline{u}(\np) 
\frac{(\Kbar - m)(\Pbar-\Kbar)}{(P-K)^2-m_\pi^2} u(\nh) 
\nonumber\\
Q\cdot{\cal B}^{(V)}
&=&            +\frac{f^2}{V^2m_\pi^2} F_1^{(V)} \frac{m}{\sqrt{E_\np E_\nh}}
 \sum_{\nk}\frac{m}{E_\nk}
\overline{u}(\np) 
\frac{(\Kbar - m) (\Pbar-\Kbar)}{(P-K)^2-m_\pi^2}
(\tau_3+i\varepsilon_{3ab}\tau_a\tau_b) u(\nh) \nonumber.
\label{gauge}
\ea

   From these relations we learn that:

\begin{itemize}
\item In the isoscalar channel the self-energy and vertex contributions 
cancel
\be
Q\cdot{\cal H}^{(S)}_p + Q\cdot{\cal B}^{(S)}=
Q\cdot{\cal H}^{(S)}_h + Q\cdot{\cal F}^{(S)} = 0 .
\ee
This is in contrast with the non-relativistic result \cite{ADM90}, 
where the self-energy is by itself gauge invariant.
\item In the isovector channel we get
\ba
Q\cdot\left[{\cal H}^{(V)}_p + {\cal B}^{(V)} \right]&=& 
\frac{2f^2}{V^2m_\pi^2} F_1^{(V)}i\varepsilon_{3ab}
\frac{m}{\sqrt{E_\np E_\nh}}\sum_{\nk}\frac{m}{E_\nk}
\overline{u}(\np) 
\frac{(\Kbar - m)(\Pbar-\Kbar)\tau_a\tau_b}{(P-K)^2-m_\pi^2}
u(\nh)
\nonumber\\
Q\cdot\left[{\cal H}^{(V)}_h + {\cal F}^{(V)}\right] &=&  
\frac{2 f^2}{ V^2m_\pi^2} F_1^{(V)}i\varepsilon_{3ab}
\frac{m}{\sqrt{E_\np E_\nh}}\sum_{\nk}\frac{m}{E_\nk}
\overline{u}(\np) 
\frac{(\Kbar-\Hslash)(\Kbar - m)\tau_a\tau_b}{(K-H)^2-m_\pi^2}
u(\nh).
\nonumber\ea
\end{itemize}
These expressions, using the Dirac equations $\Hslash u(\nh)=m u(\nh)$ and
$\overline{u}(\np) \Pbar = m \overline{u}(\np)$, can be
further simplified to yield the following four-divergence of the correlation 
current
\ba
& & \langle Q\cdot {\hat j}^C\rangle =
\frac{1}{2}Q\cdot\left[{\cal H}^{(V)}_p +{\cal B}^{(V)}+
{\cal H}^{(V)}_h + {\cal F}^{(V)}\right]
\nonumber \\
&=& 
\frac{2f^2}{V^2m_\pi^2} F_1^{(V)}i\varepsilon_{3ab}
\frac{m}{\sqrt{E_\np E_\nh}}\sum_{\nk}\frac{m}{E_\nk}
\overline{u}(\np) \tau_a \left\{
\frac{K\cdot P-m^2}{(P-K)^2-m_\pi^2} - \frac{K\cdot H-m^2}{(K-H)^2-m_\pi^2}
\right\} \tau_b u(\nh).
\nonumber\ea
This contribution is exactly cancelled by that of the MEC (seagull and
pion-in-flight) as we illustrate in what follows.

\begin{itemize}
\item
\underline{MEC}
\end{itemize}

\noindent
Using the expressions given in Section~\ref{sec:phme} 
for the ph matrix elements 
corresponding to the seagull and pion-in-flight currents
in eqs.~(\ref{Sph},\ref{Pph}), the associated four-divergences are found to be
\ba
\langle Q\cdot{{\hat j}^S}\rangle 
&=&-\frac{f^2}{V^2m_\pi^2} F_1^{(V)} i\varepsilon_{3ab} 
\frac{m}{\sqrt{E_\np E_\nh}}
\nonumber\\
&\times& \sum_{\nk}\frac{m}{E_\nk}
              \overline{u}(\np)\tau_a\tau_b
\left\{ \frac{(\Kbar-m)\Qbar}{(P-K)^2-m_\pi^2}
      + \frac{\Qbar (\Kbar-m)}{(K-H)^2-m_\pi^2} 
\right\}u(\nh) \nonumber
\\
\langle Q\cdot{{\hat j}^P}\rangle
&=&
\frac{2mf^2}{V^2m_\pi^2} F_1^{(V)} i\varepsilon_{3ab} 
\frac{m}{\sqrt{E_\np E_\nh}}
\nonumber\\
&\times& \sum_{\nk}\frac{m}{E_\nk}
              \frac{(Q^2+2H\cdot Q-2K\cdot Q)}{[(P-K)^2-m_\pi^2]
                                               [(K-H)^2-m_\pi^2]}
              \overline{u}(\np)\tau_a(\Kbar-m)
              \tau_b u(\nh). \nonumber
\ea
Exploiting the Dirac equation and after some algebra the above 
can be recast as
follows
\ba
& &\langle Q\cdot{{\hat j}^{MEC}}\rangle = \langle Q\cdot{{\hat j}^S}\rangle+
\langle Q\cdot{{\hat j}^P}\rangle
\nonumber\\
&=& - \frac{2f^2}{V^2m_\pi^2} F_1^{(V)} i\varepsilon_{3ab} 
\frac{m}{\sqrt{E_\np E_\nh}}\sum_{\nk}
\frac{m}{E_\nk}              \overline{u}(\np)\tau_a\left\{
\frac{K\cdot P-m^2}{(P-K)^2-m_\pi^2}-\frac{K\cdot H-m^2}{(K-H)^2-m_\pi^2} 
\right\}
              \tau_b u(\nh) .
\nonumber
\ea

We have thus proven that the correlation and MEC ph matrix elements satisfy 
current conservation, {\it i.e.,}
$\langle Q\cdot{{\hat j}^{C}}\rangle+
\langle Q\cdot{{\hat j}^{MEC}}\rangle = 0$.

\subsection*{Appendix B. Poles in the vertex correlation diagrams}

\begin{itemize}
\item
\underline{Forward Diagram}
\end{itemize}

\noindent
As shown in Section~\ref{sec:phme}, the forward-going diagram 
(Fig.~2d) of the v.c. ph matrix element involves
integrals of the type
\begin{equation}
\int d\nk \frac{f(\nk)}{(K+Q)^2-m^2+i\epsilon} \ .
\end{equation}
The question then is if, for given $q$ and $\omega$,
a four-momentum $K^\mu$, 
with $k<k_F$, exists such that $P'=K+Q$ is on-shell.
The three-momentum $\nk$ lies on the sphere $A$ with radius $k$ (see 
Fig. 21).
On the other hand the three-momentum $\np'$ should lie on the sphere $C$
of radius
\begin{equation}
p'=\sqrt{(E_\nk+\omega)^2-m^2} \ ,
\end{equation}
where $E_\nk=\sqrt{k^2+m^2}$ is the energy of the hole.
Moreover, since $\np'=\nk+\nq$, it
should also lie on the sphere $B$ which corresponds to $A$ translated by
an amount $\nq$.
The pole is clearly hit when the two spheres $B$ and $C$ intersect.

\begin{figure}[t]
\begin{center}
\leavevmode
\def\epsfsize#1#2{0.6#1}
\epsfbox[50 140 550 730]{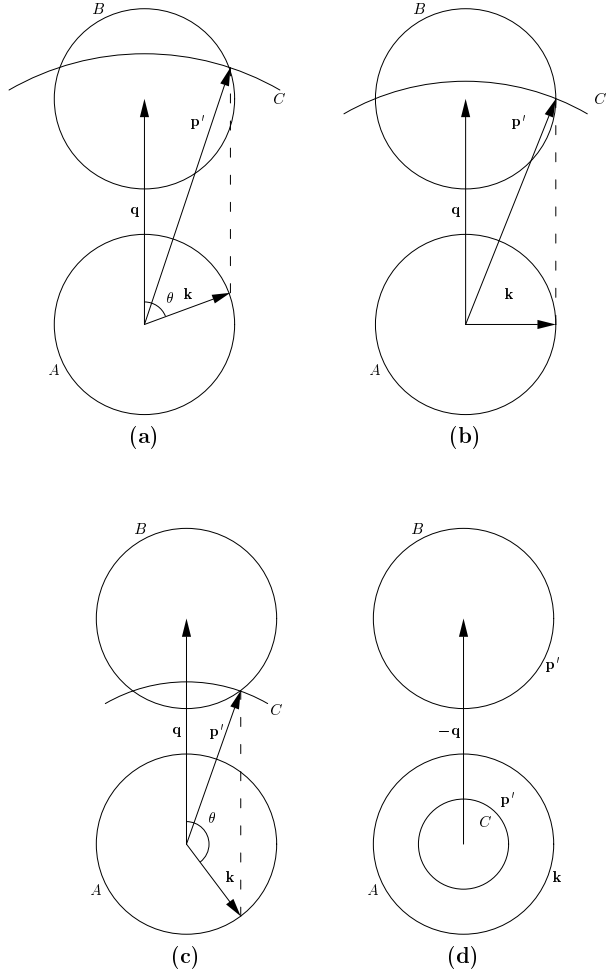}
\end{center}
\caption{
Poles in vertex correlation. Diagrams (a-c) refer to the
forward-going situation and (d) to the backward-going one.}
\end{figure}


A quantitative assessment of the position of the poles can be more easily
derived in the non-relativistic framework. In this case indeed the particle
momentum reads
\begin{equation}\label{pprima}
p'{}^2 = k^2 + 2m\omega \ .
\end{equation}
Let us now keep $q$ and $k$ fixed and study the
behaviour of the poles as a function of $\omega$. 
On the basis of eq.~(\ref{pprima}) we can consider three situations:
\begin{itemize}

\item i) $\omega=q^2/2m$ (QEP). Here $p'{}^2 = k^2+q^2$
and consequently $\nk$ is perpendicular to $\nq$, so the pole appears 
for $\cos\theta_k=0$. This situation is illustrated in Fig.~21(b).

\item ii) $\omega > q^2/2m$ (above QEP). The particle momentum satisfies
$p'{}^2 > k^2+q^2$ (Fig.~21(a)). In this situation the radius of the
sphere $C$ is larger than in the previous case (Fig.~21(a)), and therefore
$\theta_k<\frac{\pi}{2}$ and $\cos\theta_k >0$.

\item iii) $\omega < q^2/2m$ (below QEP), {\it i.e.,}
$p'{}^2 < k^2+q^2$ (Fig.~21(c)). Here 
$\theta_k>\frac{\pi}{2}$ and $\cos\theta_k <0$.

\end{itemize}


To estimate the minimum value of the hole momentum $k$ allowing for the
existence of a pole, from Figs.~21(a) and 21(c) it clearly appears
that, for given $q$ and $\omega$,
there are no poles if the value of $k$ is very small: in fact in this case
the spheres $B$ and $C$ do not intersect. More precisely, 
the range of allowed values of $k$ is limited by the condition (for $k<q$):
\begin{equation}
q-k < p' < q+k \ .
\end{equation}
Within the non-relativistic framework (see eq.~(\ref{pprima})), this condition reads
\begin{equation}
|m\Delta\omega| < qk \ ,
\end{equation}
where $\Delta\omega$ is defined as the $\omega$-shift from the QEP:
$\Delta\omega \equiv \omega-\frac{q^2}{2m}$.
Then the minumum value of $k$ allowing for the existence of a pole is
\begin{equation}
k_{\rm min} = \frac{m}{q}\left|\omega-\frac{q^2}{2m}\right|
=m\left|\frac{\lambda}{\kappa}-\kappa\right|.
\end{equation}
It then follows that at the QEP one has
$k_{\rm min}=0$ and therefore the pole always comes into play. 

\begin{itemize}
\item
\underline{Backward Diagram}
\end{itemize}

\noindent
In the case of the backward-going v.c. (Fig.~21(d)) 
the integral to be considered is
\begin{equation}
\int d\nk \frac{g(\nk)}{(K-Q)^2-m^2+i\epsilon}\ .\label{eq88}
\end{equation}
Also here one can study the conditions that allow a particle 
with four-momenutm
$P'\equiv K-Q$ to be on-shell for given $q$ and $\omega$. 
Now clearly the energy associated with $p'$ is  
$E_\np' = E_\nk -\omega$ with $E_\nk=\sqrt{k^2+m^2}$. 
The momentum $p'$ must lie on the sphere $C$ of radius
\begin{equation}
p'= \sqrt{(E_\nk-\omega)^2-m^2} <k < k_F \ ,
\end{equation}
which now is contained inside the sphere $A$ (see Fig.~21(d)).
On the other hand, since $\np'=\nk-\nq$, it
must also lie on the sphere $B$ obtained for $A$ by translating by an
amount $-\nq$. 
For $q>2k_F$ (no Pauli-blocking) the intersection between $B$ 
and $C$ vanishes:
hence no poles can be found.

\subsection*{Appendix C. Polarization propagator with nucleon self-energy}

In this appendix we derive the general expression for the self-energy
polarization propagator with any number of insertions, $\Sigma(P)$, in
the particle and/or in the hole lines. From the obtained expression
one can derive, as particular cases, the leading-order response (no
interaction lines) and the first-order self-energy response (with one
interaction line).

Let us then consider the polarization propagator
containing $n$ self-energy insertions $\Sigma(H)$ 
in the hole line and $l$ insertions $\Sigma(P)$ in the particle line. 
It is given by
\begin{equation}
\Pi^{\mu\nu}
=-i {\rm Tr}\int\frac{dh_0 d^3h}{(2\pi)^4}
\Gamma^{\mu}(Q)[S(H)\Sigma(H)]^nS(H)\Gamma^{\nu}(-Q)
[S(H+Q)\Sigma(H+Q)]^l S(H+Q).
\label{C1}
\end{equation}
Using the nucleon propagator in the medium in eq.~(\ref{eq45})
we can express the product of $n+1$ propagators appearing in eq.~(\ref{C1}) 
as a derivative of order $n$ according to
\begin{eqnarray}
[S(H)\Sigma(H)]^n S(H) 
&=&
\nonumber\\
&&
\kern -8em
=
[(\Hslash+m)\Sigma(H)]^n
(\Hslash+m)
      \left[ \frac{\theta(h-k_F)}{(H^2-m^2+i\epsilon)^{n+1}}
            +\frac{\theta(k_F-h)}{(H^2-m^2-i\epsilon h_0)^{n+1}}
      \right]
\nonumber\\
&&
\kern -8em
=
[(\Hslash+m)\Sigma(H)]^n
(\Hslash+m)
\frac{1}{n!}
\left.\frac{d^n}{d\alpha^n}\right|_{\alpha=0}
      \left[ \frac{\theta(h-k_F)}{H^2-\alpha-m^2+i\epsilon}
            +\frac{\theta(k_F-h)}{H^2-\alpha-m^2-i\epsilon h_0}
      \right]
\nonumber\\
&&
\kern -8em
=
[(\Hslash+m)\Sigma(H)]^n
(\Hslash+m)
\nonumber\\
&&
\kern -8em
\times
\frac{1}{n!}
\left.\frac{d^n}{d\alpha^n}\right|_{\alpha=0}
\left[
       \frac{1}{H^2-\alpha-m^2+i\epsilon}
      + 2\pi i
        \theta(k_F-h)
        \delta(H^2-\alpha-m^2)\theta(h_0) 
\right].
\end{eqnarray}
A similar equation holds for the propagation of a particle.
The global polarization propagator, with both the particle and the
hole line dressed, can then be written as
\begin{eqnarray}
\Pi^{\mu\nu}
&=&
-i \frac{1}{n!l!}
\left.\frac{d^n}{d\alpha^n}\right|_{\alpha=0}
\left.\frac{d^l}{d\beta^l}\right|_{\beta=0}
\nonumber\\
&&
\mbox{}\times
{\rm Tr}\int\frac{dh_0 d^3h}{(2\pi)^4}
\Gamma^{\mu}(Q)[(\Hslash+m)\Sigma(H)]^n(\Hslash+m)\Gamma^{\nu}(-q)
[(\Pbar+m)\Sigma(P)]^l (\Pbar+m)
\nonumber\\
&&\mbox{}\times
\left[
       \frac{1}{H^2-\alpha-m^2+i\epsilon}
      + 2\pi i
        \theta(k_F-h)
        \delta(H^2-\alpha-m^2)\theta(h_0) 
\right]
\nonumber\\
&&\mbox{}\times
\left[
       \frac{1}{P^2-\beta-m^2+i\epsilon}
      + 2\pi i
        \theta(k_F-p)
        \delta(P^2-\beta-m^2)\theta(p_0) 
\right],
\end{eqnarray}
where $P=H+Q$. 

It gives rise to four terms. One of these contains the product of the
two free propagators, namely
\begin{equation}
       \left(\frac{1}{H^2-\alpha-m^2+i\epsilon}\right)
       \left(\frac{1}{P^2-\beta-m^2+i\epsilon}\right) \ ,
\end{equation}
and yields the divergent vacuum contribution, which should be
subracted out and pertains to a domain beyond nuclear physics. 
Then, after some algebra, one obtains
\begin{eqnarray}
\Pi^{\mu\nu}
-\Pi^{\mu\nu}_0
&=&
2\pi
\left.\frac{d^n}{d\alpha^n}\right|_{\alpha=0}
\left.\frac{d^l}{d\beta^l}\right|_{\beta=0}
\int\frac{dh_0 d^3h}{(2\pi)^4}
I^{\mu\nu}_{nl}(H,P,Q)
\nonumber\\
&&
\kern -4em 
\mbox{}\times
\left[
\rule[-7mm]{0mm}{14mm}
       \frac{\theta(k_F-p)
             \delta(P^2-\beta-m^2)\theta(p_0) 
            }{H^2-\alpha-m^2+i\epsilon}
+
        \frac{\theta(k_F-h)
             \delta(H^2-\alpha-m^2)\theta(h_0) 
            }{H^2-\beta-m^2+i\epsilon}
\right.
\nonumber\\
&&
\kern -4em
\mbox{}+
\left.
\rule[-7mm]{0mm}{14mm}
        2\pi i 
        \theta(k_F-p)
        \theta(k_F-h)
        \delta(P^2-\beta-m^2)
        \delta(H^2-\alpha-m^2)
        \theta(P_0) \theta(H_0) 
\right],
\end{eqnarray}
where the function
\begin{eqnarray}
& &I^{\mu\nu}_{nl}(H,P,Q)=
I^{\mu\nu}_{nl}(h_0,\nh;p_0,\np;q_0,\nq)
\nonumber\\
&\equiv&
\frac{1}{n!l!}
{\rm Tr}
\left[
\Gamma^{\mu}(Q)[(\Hslash+m)\Sigma(H)]^n(\Hslash+m)\Gamma^{\nu}(-Q)
[(\Pbar+m)\Sigma(P)]^l (\Pbar+m)
\right]
\end{eqnarray}
has been introduced.
According to eq.~(\ref{eq4}) we then obtain for the nuclear tensor
\begin{eqnarray}
& &-\frac{V}{\pi} {\rm Im}\,
\left[\Pi^{\mu\nu}-\Pi^{\mu\nu}_0\right] \nonumber \\
&=& 2\pi
\left.\frac{d^n}{d\alpha^n}\right|_{\alpha=0}
\left.\frac{d^l}{d\beta^l}\right|_{\beta=0}
\int\frac{dh_0 d^3h}{(2\pi)^4}
I^{\mu\nu}_{nl}(H,P,Q)
        \delta(P^2-\beta-m^2)
        \delta(H^2-\alpha-m^2)
\nonumber\\
&&\mbox{}\times
\left[
        \theta(k_F-p)
        \theta(p_0)
+
        \theta(k_F-h)
        \theta(h_0) 
-2
        \theta(k_F-p)
        \theta(k_F-h)
        \theta(p_0)
        \theta(h_0) 
\right].
\nonumber\\
\end{eqnarray}

Exploiting now the identity
\begin{eqnarray}
\left[
        \theta(k_F-p)
        \theta(p_0)
+
        \theta(k_F-h)
        \theta(h_0) 
-2
        \theta(k_F-p)
        \theta(k_F-h)
        \theta(p_0)
        \theta(h_0) 
\right]
\nonumber\\
\mbox{}=
        \theta(k_F-h)
        \theta(h_0) 
[1-\theta(k_F-p)\theta(p_0)]
+
        \theta(k_F-p)\theta(p_0)
[1-     \theta(k_F-h)\theta(h_0)] 
\end{eqnarray}
we can recast the above according to
\begin{equation}
-\frac{1}{\pi} {\rm Im}\, [\Pi^{\mu\nu} -\Pi^{\mu\nu}_0]
= W^{\mu\nu}_+ + W^{\mu\nu}_- \ ,
\end{equation}
where 
\begin{eqnarray}
W^{\mu\nu}_+(Q) 
&=&
2\pi
\left.\frac{d^n}{d\alpha^n}\right|_{\alpha=0}
\left.\frac{d^l}{d\beta^l}\right|_{\beta=0}
\int\frac{dh_0 d^3h}{(2\pi)^4}
I^{\mu\nu}_{nl}(H,P,Q)
        \delta(P^2-\beta-m^2)
        \delta(H^2-\alpha-m^2)
\nonumber\\
&&
\kern 3cm
\mbox{}\times
        \theta(k_F-h)
        \theta(h_0) 
[1-\theta(k_F-p)\theta(p_0)]
\label{wmunuplus}
\end{eqnarray}
and
\begin{eqnarray}
W^{\mu\nu}_-(Q)
&=&
2\pi
\left.\frac{d^n}{d\alpha^n}\right|_{\alpha=0}
\left.\frac{d^l}{d\beta^l}\right|_{\beta=0}
\int\frac{dh_0 d^3h}{(2\pi)^4}
I^{\mu\nu}_{nl}(H,P,Q)
        \delta(P^2-\beta-m^2)
        \delta(H^2-\alpha-m^2)
\nonumber\\
&&
\kern 4cm
\mbox{}\times
        \theta(k_F-p)
        \theta(p_0) 
[1-\theta(k_F-h)\theta(h_0)] \ .
\label{wmunuminus}\end{eqnarray}
Now eq.~(\ref{wmunuplus}) corresponds to the usual hadronic tensor for electron 
scattering, whereas eq.~(\ref{wmunuminus}) corresponds to a process with
negative energy and momentum transfer, hence it should be disregarded.

Next the integration with respect to $h_0$ can be explicitly performed
by using
the second of the $\delta$-functions appearing in eq.~(\ref{wmunuplus}). 
After some algebra one then gets for the hadronic tensor the expression

\begin{eqnarray}
W^{\mu\nu}
&=&
\left.\frac{d^n}{d\alpha^n}\right|_{\alpha=0}
\left.\frac{d^l}{d\beta^l}\right|_{\beta=0}
\int\frac{d^3h}{(2\pi)^3}
\frac{I^{\mu\nu}_{nl}(E'_\nh(\alpha),\nh;E'_\nh(\alpha)+q_0,\np;q)}
{2E'_\nh(\alpha)}
\nonumber\\
&&\times
\frac{\delta(E'_\nh(\alpha)+q_0-E'_\np(\beta))}{2E'_\np(\beta)}
        \theta(k_F-h)
        \theta(p-k_F) \ ,
\end{eqnarray}
where 
\ba
E'_\nh(\alpha) &=& \sqrt{h^2+\alpha+m^2}\\
E'_\np(\beta) &=& \sqrt{p^2+\beta+m^2}.
\ea

Finally, exploiting the first $\delta$-function in eq.~(\ref{wmunuplus}),
one can cast the response function in the form
\begin{eqnarray}
W^{\mu\nu}
&=&
\left.\frac{d^n}{d\alpha^n}\right|_{\alpha=0}
\left.\frac{d^l}{d\beta^l}\right|_{\beta=0}
\int\frac{d^3h}{(2\pi)^3}
\frac{I^{\mu\nu}_{nl}(E'_\nh(\alpha),\nh;E'_\np(\beta),\np;q)}
{4E'_h(\alpha)E'_\np(\beta)}
\nonumber\\
&&\times
\delta(E'_\nh(\alpha)+q_0-E'_\np(\beta))
        \theta(k_F-h)
        \theta(p-k_F).
\label{Wnl}
\end{eqnarray}
For $n=l=0$ one gets the well known free hadronic tensor
\be
W^{\mu\nu}=
\int\frac{d^3h}{(2\pi)^3}
\frac{I^{\mu\nu}_{00}(E_\nh,\nh;E_\np,\np;q)}
{4E_\nh E_\np}
\delta(E_\nh+q_0-E_\np)
        \theta(k_F-h)
        \theta(p-k_F) \ ,
\ee
for $n=1$, $l=0$ the first-order hole self-energy hadronic tensor

\be
W^{\mu\nu}=
\left.\frac{d}{d\alpha}\right|_{\alpha=0}
\int\frac{d^3h}{(2\pi)^3}
\frac{I^{\mu\nu}_{10}(E'_\nh(\alpha),\nh;E_\np,\np;q)}
{4E'_\nh(\alpha)E_\np}
\delta(E'_\nh(\alpha)+q_0-E_\np)
        \theta(k_F-h)
        \theta(p-k_F) \label{eq106}
\ee
and finally for $n=0$, $l=1$ the first-order particle self-energy hadronic
tensor
\be
W^{\mu\nu}=
\left.\frac{d}{d\beta}\right|_{\beta=0}
\int\frac{d^3h}{(2\pi)^3}
\frac{I^{\mu\nu}_{01}(E_\nh,\nh;E'_\np(\beta),\np;q)}
{4E_\nh E'_\np(\beta)}
\delta(E_\nh+q_0-E'_\np(\beta))
        \theta(k_F-h)
        \theta(p-k_F).
\label{eq107}
\ee
Note that the last two expressions coincide, as they should, with the 
ones already given in Section~\ref{sec:tensor}.

The hadronic tensor in eq.~(\ref{Wnl}) is one of the $(n+l)$-th order 
contributions
to the full Hartree-Fock polarization propagator, which should be obtained
as an infinite sum of all perturbative orders. 
It should be stressed that in a basically non-relativistic framework
the HF series can be summed (see, e.g., \cite{Barb93}), 
whereas in our fully relativistic framework the summation 
is far more difficult to perform because of the matrix
nature of the self-energy $\Sigma$~\cite{Serot}.


\end{document}